\documentclass[12pt]{article}
\pdfoutput=1 

\usepackage{jheppub} 
                                          
\usepackage{placeins}
\usepackage{tabularx}
\usepackage{amsmath}

\newcommand{\rmii}[1]{{\mbox{\tiny\rm{#1}}}}
\newcommand{\beq}{\begin{equation}}
\newcommand{\eeq}{\end{equation}}
\newcommand{\bea}{\begin{align}}
\newcommand{\eea}{\end{align}}
\newcommand{\beas}{\begin{align*}}
\newcommand{\eeas}{\end{align*}}

\newcommand{\bp}{{\bf p}}

\newcommand{\eps}{{\varepsilon}}
\newcommand{\mD}{m_\rmii{D}}

\newcommand{\vE}{\vec E}
\newcommand{\rh}{\hat r}
\newcommand{\vr}{\vec r}
\newcommand{\vp}{\vec p}

\newcommand{\Tint}[1]{{\hbox{$\sum$}\!\!\!\!\!\!\!\int\,}_{\!\!\!\!\raise-0.9ex\hbox{$\scriptstyle{#1}$}}}
\renewcommand{\Re}{\rm Re}
\renewcommand{\Im}{\rm Im}

\definecolor{dunkelgrau}{rgb}{0.43,0.43,0.43}

\title{\bf\large Quarkonium at finite temperature: Towards realistic phenomenology from first principles}

\author[a]{Yannis Burnier,}
\author[b]{Olaf Kaczmarek}
\author[c]{and Alexander Rothkopf}

\affiliation[a]{Institute of Theoretical Physics, EPFL, CH-1015 Lausanne, Switzerland}
\affiliation[b]{Fakult\"at f\"ur Physik, Universit\"at Bielefeld, D-33615 Bielefeld, Germany}
\affiliation[c]{Institute for Theoretical Physics,  Heidelberg
  University, Philosophenweg 16, 69120 Heidelberg, Germany}

\emailAdd{yannis.burnier@epfl.ch}
\emailAdd{okacz@physik.uni-bielefeld.de}
\emailAdd{rothkopf@thphys.uni-heidelberg.de}

\date{\today}

\abstract{We present the finite temperature spectra of both bottomonium and charmonium, obtained from a consistent lattice QCD based potential picture. Starting point is the complex in-medium potential extracted on full QCD lattices with dynamical u,d and s quarks, generated by the HotQCD collaboration. Using the generalized Gauss law approach, vetted in a previous study on quenched QCD, we fit ${\rm Re}[V]$ with a single temperature dependent parameter $m_D$, the Debye screening mass, and confirm the up to now tentative values of ${\rm Im}[V]$. The obtained analytic expression for the complex potential allows us to compute quarkonium spectral functions by solving an appropriate Schr\"odinger equation. These spectra exhibit thermal widths, which are free from the resolution artifacts that plague direct reconstructions from Euclidean correlators using Bayesian methods. In the present adiabatic setting, we find clear evidence for sequential melting and derive melting temperatures for the different bound states. Quarkonium is gradually weakened by both screening (${\rm Re}[V]$) and scattering (${\rm Im}[V]$) effects that in combination lead to a shift of their in-medium spectral features to smaller frequencies, contrary to the mass gain of elementary particles at finite temperature.
}

\begin{document}

\maketitle


\flushbottom

\FloatBarrier

\section{Introduction}

Heavy quarkonium, the bound states of a heavy quark and anti-quark ($c\bar{c}$ or $b\bar{b}$), are a unique tool to simplify the complexity inherent in the physics of the strong interactions. Instead of having to deploy the full force of quantum field theory, we may consider a Schr\"odinger equation with an effective potential to describe the dynamics of these bound states in vacuum and in-medium. At zero temperature the two main characteristics of QCD, asymptotic freedom and confinement manifest themselves clearly in the form of a Coulombic behavior with a running coupling at small distances and a non-perturbative linear rise at large distances \cite{Koma:2006si}. I.e. we can learn about key features of the strong interactions by inspecting this simple potential. Vice versa, from the knowledge of the potential, we can reproduce even quantitatively a significant part of vacuum quarkonium physics (e.g. the bound state spectrum below the open heavy quark threshold \cite{Quigg:1979vr,Eichten:1995ch}).

The question of how such a potential arises from the microscopic theory of QCD has been answered in detail by advances made in the research on the effective field theories (EFT) NRQCD and pNRQCD. The latter has been introduced at first at $T=0$ in \cite{Brambilla:1999xf} and investigated in the non-perturbative regime in \cite{Brambilla:2000gk,Pineda:2000sz}, with both cases being reviewed in detail in \cite{Brambilla:2004jw}. The generalization to finite temperature in a perturbative setting followed in \cite{Brambilla:2008cx}.  

In the presence of a separation of scales, the combination of NRQCD and pNRQCD offers a systematic power counting prescription in the heavy quark velocity $v$, to setup a simplified description of the bound state evolution at \textit{soft} energy scales ($E_{\rm soft}\sim m_Q v$) in terms of singlet and color octet wavefunctions. I.e. we do not have to explicitly describe the physics at the much higher \textit{hard} scale ($E_{\rm hard}\sim 2m_Q$)\footnote{The strength of the EFT approach lies in the fact that any residual influence of the physics of the hard scale can be systematically included via the matching of Wilson coefficients, which goes beyond the ability of direct methods, such as e.g. the historic $T=0$ Wilson loop approach \cite{Barchielli:1986zs}}.  And indeed the heavy quark rest mass (in this work we use $m_c\simeq 1.472$GeV and $m_b=4.882$GeV) and the characteristic scale of quantum fluctuations in QCD, usually denoted by $\Lambda_{\rm QCD}\simeq 200$MeV are far enough apart to warrant a quantitatively reliable potential description in vacuum. 

Effective field theory has furthermore reminded us that the potential between two infinitely heavy quarks is an inherently Minkowski-time quantity. Starting out from fundamental QCD, the evolution of a $Q\bar{Q}$ can be described by the thermally averaged unequal-time point-split meson-meson correlator
\begin{eqnarray}
D^>({\bf r},t)
=\Big\langle M({\bf x},{\bf y},t) M^\dagger({\bf x}_0,{\bf y}_0,0) \Big\rangle , 
\label{Eq:ForwProp}
\end{eqnarray}
with the meson operator defined as
$M({\bf x},{\bf y},t)=\bar{Q}({\bf x},t)\gamma^\mu U[{\bf x},{\bf y}]Q({\bf y},t)$. $\gamma^\mu$ refers to the Dirac matrices and $U[{\bf x},{\bf y}]$ denotes a straight Wilson line connecting $({\bf x},t)$ and $({\bf y},t)$.  The relative coordinate  enters as  ${\bf r}={\bf x}-{\bf y}$. 
  
In the limit of infinite quark mass, where the static constituents are truly test particles, \eqref{Eq:ForwProp} can be identified with the medium averaged unequal time correlation function of quarkonium singlet wavefunctions in the language of the EFT. It is this wavefunction correlator for which a Schroedinger equation and thus an in-medium potential will be established. For static quarks the spatial separation $r=|{\bf r}|$ becomes an external parameter of the theory and their evolution traces out a rectangle over time. Formally it has been shown that eq.\eqref{Eq:ForwProp} reduces to the rectangular real-time Wilson loop \cite{Brambilla:2004jw}
\beq
W_\square(r,t)=\Big\langle {\rm Tr} \Big( {\rm exp}\Big[-ig\int_\square dx^\mu A_\mu^aT^a\Big] \Big) \Big\rangle\label{Eq:WilsonLoopDef}.
\eeq
As such, this object still contains the physics of all scales: hard, soft and ultra-soft, i.e. in general it fulfills an equation of motion \cite{Laine:2007qy}
\beq
i\partial_tW_\square(r,t)=\Phi(r,t)W_\square(r,t)
\eeq
with $\Phi(r,t)$ being a time- and space dependent complex function. 

In case that a potential picture for the static $Q\bar{Q}$ system is applicable, the function $\Phi(r,t)$ has to asymptote to a time independent function $V(r)$. In turn it will dominate the evolution of $W_\square(r,t)$ at times much larger than the intrinsic scales of e.g. the gluons  and light quarks in the medium. Formally we may then write
\beq
V(r)=\lim_{t\to\infty} \frac{i\partial_t W(t,r)}{W(t,r)}\label{Eq:VRealTimeDef}.
\eeq
This limit reflects the fact that in order to replace a retarded, i.e. gluon mediated interaction with an instantaneous potential, the gluons must have been exchanged between the heavy quarks multiple times.

In the presence of the additional scale $T$ we need to ascertain how reliable a description of realistic, i.e. finite-mass, quarkonium is in terms of the static potential alone. Two forms of so called relativistic corrections can adversely affect the accuracy of the lowest order approximation. The first kind remains fully within the potential picture, i.e. finite mass corrections to the static potential can become significant, as they are proportional to the relative velocity of the heavy quarkonium system. These corrections can be systematically computed (see e.g. \cite{Brambilla:2000gk,Pineda:2000sz,Koma:2006si}), in vacuum they are found to be small and we expect the same to be true at finite temperature. Their non-perturbative determination from finite temperature lattice QCD will be the aim of a future study. 

The other contributions are so called non-potential effects, arising from physics that cannot be recast into the form of a time independent term in the Schr\"odinger equation. One way they can enter in the perturbative formulation of effective field theories if the ultra-soft gluons are still kept as explicit degrees of freedom (see e.g. \cite{Brambilla:2004jw,Bazavov:2014soa}). In a non-perturbative setting, if non-potential effects become sizable, the late time evolution of the Wilson loop cannot be described by a simple time-independent potential $V(r)$. As will be discussed in the next section we find that at the temperatures and inter-quark distance investigated here, the size of such contributions remains insignificant.

Interestingly the definition of the potential from the Wilson loop in real-time coincides with the late $\tau$ limit in Euclidean time in the case of $T=0$. This constitutes the basis for the successful extraction of the static zero temperature potential from lattice QCD simulations, which are carried out solely in an imaginary time setting. At finite temperature, the real-time definition of the potential does not change, however the imaginary time axis becomes compactified and its finite extend encodes vital physics information, i.e. the inverse temperature. Hence the straight forward connection between the late Minkowski time limit and the Euclidean Wilson loop at maximum $\tau=\beta$ is absent.

For more than two decades, the absence of an effective-field theory based definition for the the in-medium potential has led theorists to embrace model potentials that were defined directly from Euclidean time observables \cite{Nadkarni:1986cz}, readily calculable in lattice QCD. In particular two quantities have gained popularity as model potentials, the color singlet free energies in Coulomb gauge
\beq
e^{\beta F^{(1)}(r)}=\Big\langle {\rm Tr}\Big[ \Omega(r)\Omega^\dagger(0) \Big]\Big\rangle_{\rm CG},\quad \Omega(r)={\rm exp}\Big[-ig\int_0^\beta d\tau A_\mu^a(r,\tau)T^a\Big]
\eeq
defined from the correlator of Polyakov loops $\Omega(r)$ and a derived quantity the color singlet internal energies $U^{(1)}=F^{(1)}-TS$. While these quantities exhibit a behavior compatible with the expectations for e.g. Debye screening of the interaction between the heavy quarks in the deconfined phase, it could be shown that they do not match the potential for the quark anti-quark system at finite temperature \cite{Burnier:2009bk,Brambilla:2010xn}.

And indeed neither one of these fully real quantities by themselves can take the role of a static in-medium heavy quark potential, as we have learned from a ground breaking series of works starting with Laine et.~al. \cite{Laine:2007qy,Brambilla:2008cx} in 2007. In their study the real-time definition \eqref{Eq:VRealTimeDef} was evaluated in a resummed perturbative framework, called the hard-thermal loop approximation. This approach contains a gauge invariant resummation of an infinite number of Feynman diagrams and has been shown to capture many key features of QCD at high temperature reliably. As the Wilson loop in Minkowski time is an in general complex quantity, the authors observed that the potential too takes on complex values at finite temperature
\beq
V_{\rm HTL}(r)= -\tilde\alpha_s\left[\mD+\frac{e^{-\mD r}}{r}
+iT\phi(\mD r)\right]+\mathcal{O}(g^4)\label{Eq:VHTL},
\eeq
with 
\beq
\phi(x)=2 \int_0^\infty dz \frac{z}{(z^2+1)^2}\left(1-\frac{\sin(xz)}{xz}\right).\label{phi}
\eeq
Here we absorb a factor $C_F$ in the definition of the coupling constant $\tilde\alpha_s=\frac{g^2C_F}{4\pi}$ to connect to the conventions in the phenomenology literature. It was furthermore shown that the real-part of this complex potential itself does not coincide with the color-single free energies in hard-thermal loop resummed perturbation theory in \cite{Burnier:2009bk,Brambilla:2010xn}, even though the absolute deviations are comparatively small.

The fact that the in-medium potential is a complex quantity not only reflects a quantitative change but necessitates a qualitatively different perspective on the physics it describes. I.e. besides screening of the force between the heavy quarks (Debye screening) seen in the real-part, the effects of scattering of light medium degrees \cite{Laine:2007qy,Brambilla:2013dpa} with the heavy quarks (Landau damping) related to ${\rm Im}[V]$ further weaken the bound state in the QCD medium. In fact, depending on the hierarchy of scales present, the imaginary part can be related to different phenomena, such as the breakup of a color singlet to an octet configuration \cite{Brambilla:2008cx} and in turn to the dissociation of the $Q\bar{Q}$ system into gluons \cite{Brambilla:2011sg}.

Only these effects taken together can give a consistent picture of heavy quark bound states at finite temperature. I.e. a simple estimate of the dissolution of a heavy quarkonium state solely on the basis of vanishing binding energy is not sufficient. This in turn has consequences for heavy quarkonium phenomenology in general, where e.g. the melting temperatures enter as input into transport model calculations. 

We would like to stress that the in-medium potential discussed here does not directly govern the evolution of the actual wave-function of the heavy quarkonium system. By construction it instead enters in the Schr\"odinger equation of the real-time Wilson loop, which in the EFT language corresponds to the thermally averaged correlator of unequal time wavefunctions. The imaginary part of the potential hence describes the decay of this correlator over time, which does not directly imply the annihilation of the heavy-quarks. In fact, due to the non-relativistic approximation we operate under, $Q$ and $\bar{Q}$ remain in the system forever. However, even if the norm of the quarkonium wavefunction is preserved in unitary time evolution, its correlation with the initial state can still decay with time, a phenomenon known as decoherence. It is an active area of research to answer the question how to connect the complex in-medium potential to the evolution of the bound state wave-function, which has not yet been answered in the effective field theory setting of pNRQCD. The concept of open-quantum systems (see e.g. \cite{Akamatsu:2011se}) has proven insightful in this regard.

One possible way to elucidate the physics encoded in the complex potential is to compute the spectral function $\rho(\omega)$ of heavy quarkonium. This quantity is related to the heavy quark current-current correlator, which can be obtained from \eqref{Eq:ForwProp} by carefully taking the limit of vanishing point splitting. In section \ref{section3} we will hence solve an appropriate Schr\"odinger equation to compute $\rho(\omega)$. Once computed one can observe how the formerly delta-like bound state peaks present at $T=0$ broaden and shift as screening and scattering modifies the $Q\bar{Q}$ state with increasing temperature. Choosing a popular criterion of melting temperature \cite{Laine:2007qy,Kakade:2015xua} at the point where binding energy and spectral width coincide we can furthermore determine the point of dissolution for different bottomonium and charmonium states. 

Changes in spectral structure are of particular interest as they are directly related to changes in the dilepton emission from quarkonium decay. The dilepton emission rate is given by a simple product of the in-medium spectral function with the Bose-Einstein factor \cite{McLerran:1984ay}
\beq
\frac{dR_{\ell \bar \ell}}{d^4 P}=-\frac{Q_q^2\alpha_e^2}{3\pi^3P^2} n_B(p_0)\rho(P), \label{dilepton_prod}
\eeq
where $Q_q$ denotes the electric charge of the heavy quark considered in units of $e$, the four momentum is $P=(p_0,\bp)$ and the finite mass of the leptons has been neglected (for a more detailed discussion of the above formula see ref.~\cite{Laine:2013vma}). I.e. if a bound state or its remnant appears as well defined peaked feature in an in-medium spectral function, the area under such a structure informs us about the experimentally accessible dilepton emission of that state. Note that while the above relation is only applicable to the decay of a quarkonium state in a thermalized static plasma, it can nevertheless provide us with vital physics insight as we will see in sec.~\ref{sec:charmonium}.

This paper is organized as follows. We start  sec.~ \ref{section2} with a review on recent progress in the extraction of the values of the complex in-medium potential from Euclidean time lattice QCD simulations.  In order to deploy the lattice potential in phenomenological applications,  sec.~\ref{sec:GaussLaw} discusses the generalized Gauss law ansatz developed in ref.~\cite{Burnier:2015nsa}, which provides an analytic formula for $\Re[V]$ and $\Im[V]$ depending on a single temperature dependent parameter $m_D$ the Debye mass. Tuning $m_D$ we are able to reproduce the lattice values of $\Re[V]$ and confirm the yet tentative values for $\Im[V]$. These values for $m_D$ are compared to perturbation theory.
Section \ref{section3} is concerned with using the continuum corrected potential to investigate the phenomenology of in-medium heavy quarkonium. We will calculate the spectra for the bottomonium and charmonium S-wave channel and investigate their modification with increasing temperature. Besides giving the melting temperatures for different states, we will determine the $\Psi'$ to $J/\Psi$ ratio at the chiral crossover and give a rough estimate for the suppression of bottomonium  in a heavy-ion collision compared to $\rm p+p$. We close with a  conclusion in sec.~ \ref{sec:concl}.

\section{Lattice QCD potential and Debye screening mass}
\label{section2}

While the perturbative computation of the potential has contributed significantly to our understanding of the  physics involved, it is not sufficient for the description of the experimentally relevant temperature regime around the phase transition. There the quark gluon plasma can indeed be considered strongly interacting, exemplified, e.g. by the large value of the trace anomaly \cite{Bazavov:2014pvz}. This calls for an evaluation of the potential definition \eqref{Eq:VRealTimeDef} in lattice QCD, which at first seems unfeasible, since direct access to real-time quantities, such as the Wilson loop \eqref{Eq:WilsonLoopDef} is not possible. Conceptual and technical advances in the extraction of real-time information from lattice QCD simulations over the last few years however have made such an evaluation possible, as we will discuss below. 

\subsection{The complex in-medium potential from lattice QCD}

The real-time Wilson loop cannot be directly accessed in lattice QCD simulations, as they are performed in Euclidean time. One strategy, proposed in \cite{Rothkopf:2009pk} and applied for the first time in \cite{Rothkopf:2011db} to circumvent this problem is to resort to a spectral decomposition of the Wilson loop, which simply amounts to a Fourier transform over a positive definite spectral function $\rho$
\beq
 \nonumber W_\square(\tau,r)=\int d\omega e^{-\omega \tau} \rho_\square(\omega,r)\,
\leftrightarrow\, \int d\omega e^{-i\omega t} \rho_\square(\omega,r)= W_\square(t,r).
\eeq
The fact that the time dependence is explicit in the integral kernel tells us that both the real-time and Euclidean time Wilson loop are described by the same spectral function. Hence extracting the values of $\rho$ from imaginary time simulation data will give access to the real-time Wilson loop and in turn to the potential. Inserted into the defining equation \eqref{Eq:VRealTimeDef} the spectrum itself can be related to the values of the potential 
\begin{align}
V(r)=\lim_{t\to\infty}\int d\omega\, \omega e^{-i\omega t} \rho_\square(\omega,r)/\int d\omega\, e^{-i\omega t} \rho_\square(\omega,r). \label{Eq:PotSpec}
\end{align}

\begin{figure}
\centering
 \hspace{-1.2cm} \includegraphics[scale=0.28]{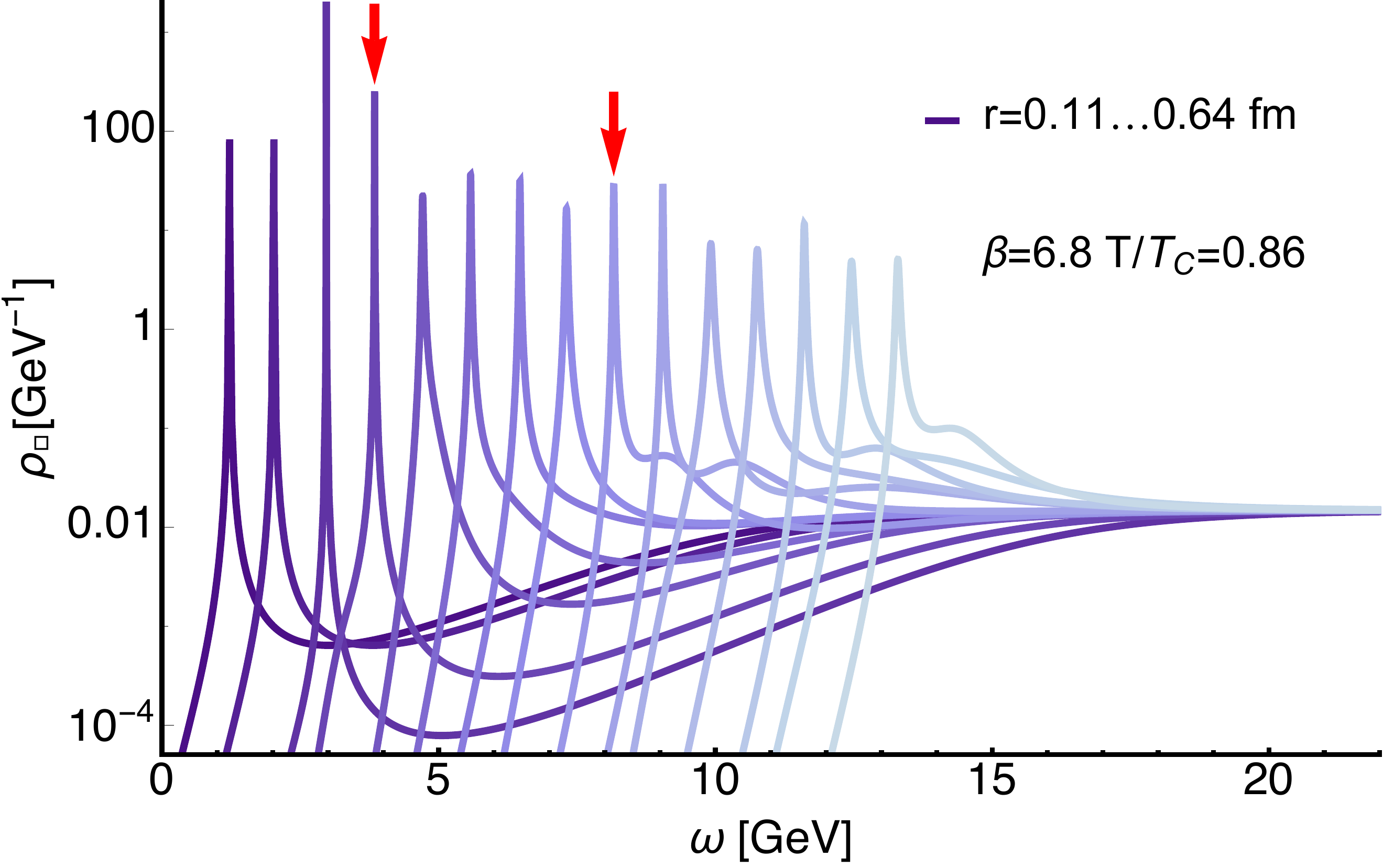}\hspace{0.4cm}
 \includegraphics[scale=0.28]{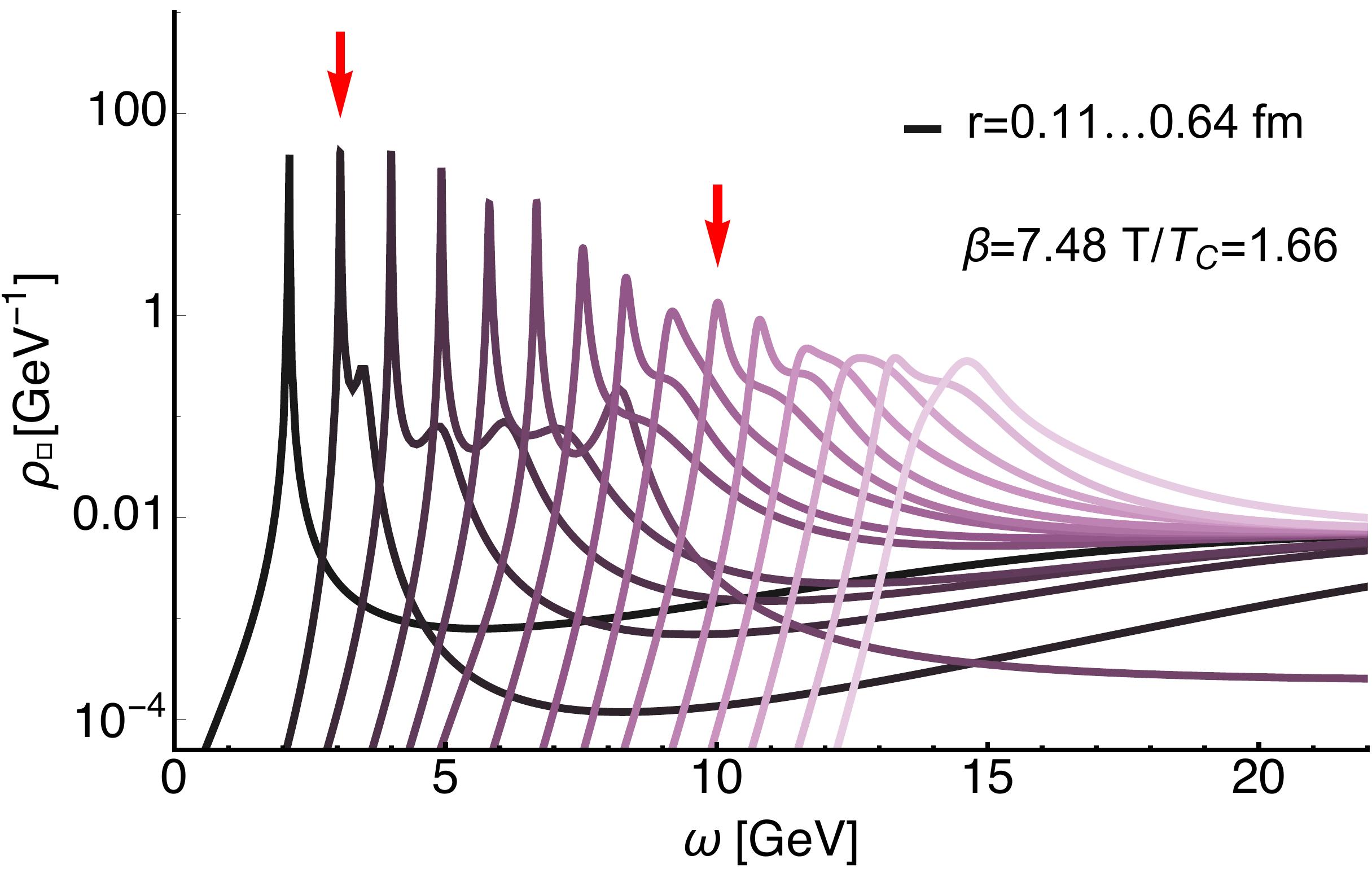}\\
   \hspace{-0.45cm}\begin{minipage}{6in}
  \centering
  \raisebox{-0.5\height}{\includegraphics[scale=0.265]{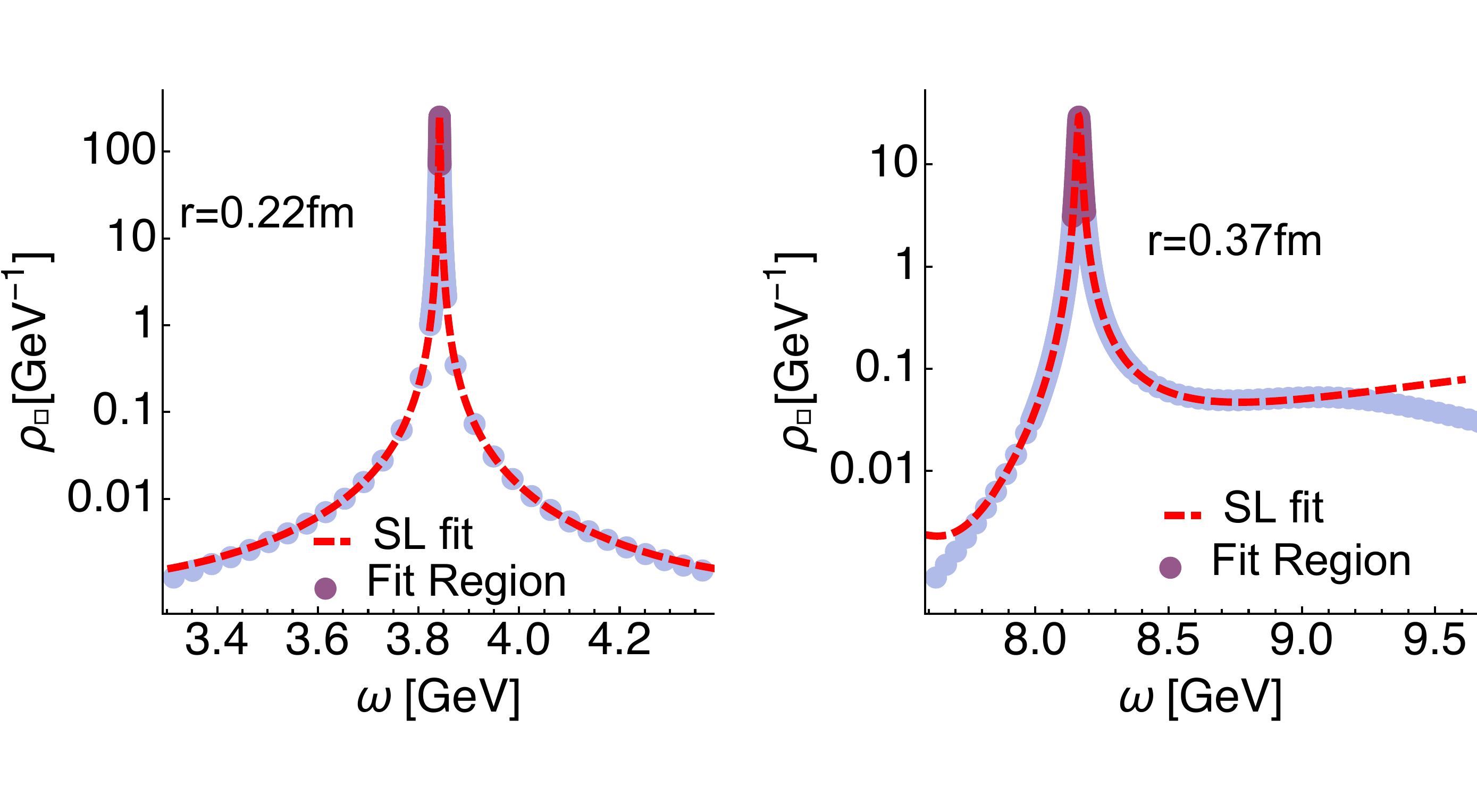}}
  \hspace*{.2in}
  \raisebox{-0.5\height}{ \includegraphics[scale=0.24]{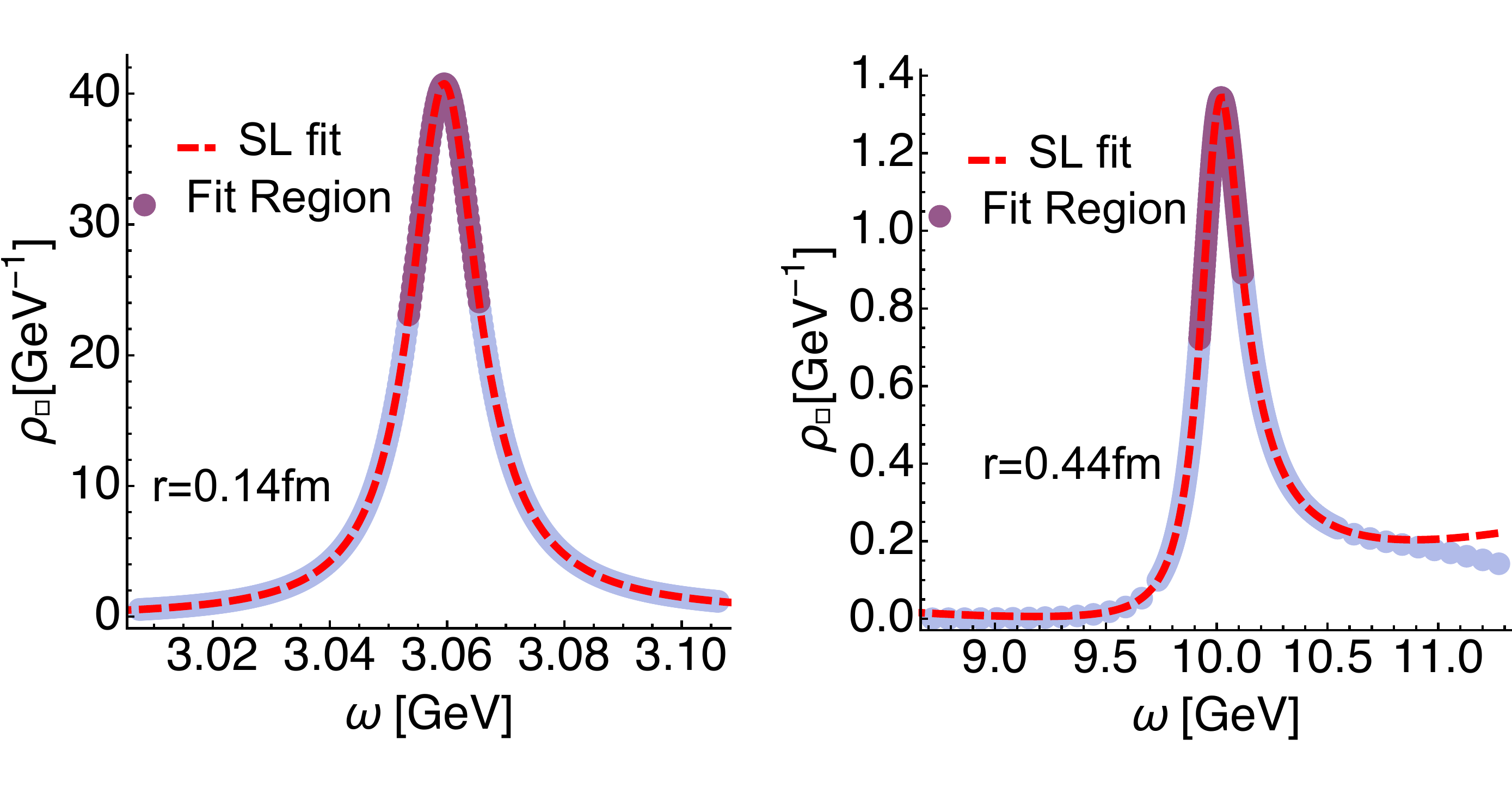}}
\end{minipage}
 \caption{(top) A selection of spectral functions from the Euclidean Wilson Line correlators in Coulomb gauge, extracted from $48^3\times12$ HotQCD asqtad lattices using a novel Bayesian reconstruction prescription \cite{Burnier:2013nla} at the lowest (left) and highest (right) temperatures. To better observe their peak and shoulder structures, individual spectra are shifted manually in frequency. For $T/T_C=0.86$ a well defined lowest lying peak is present at any of the shown distances $r$, while at $T/T_C=1.66$ the reduced physical extend degrades the reconstruction at larger $r$. (bottom) Illustration of the skewed Lorentzian (SL) nature of the individual spectral peaks. Using eq.\eqref{skewedrho} up to quadratic polynomial order, we fit the points in the dark red region close to the maximum in each case. The resulting fit functions (dashed orange) reproduce the shape of the spectra (blue circles) quantitatively far beyond the actual fitting region. Note that this indicates that the peak is a Lorentzian and not e.g. a Gaussian}\label{Fig:WlineSpec}
\end{figure}

Unfortunately extracting the spectrum from Euclidean time simulation data is an inherently ill-defined inverse problem, as one seeks to determine the form of a continuous function from a finite and noisy set of individual points. Only by using additional prior information, such as the positive definiteness of the spectrum or smoothness assumptions is it possible to give meaning to this problem, a strategy usually referred to as Bayesian inference. Established implementation of this approach, such as the Maximum Entropy Method \cite{Asakawa:2000tr} or extended MEM \cite{Rothkopf:2012vv} have been shown to fail to produce satisfactory results when deployed in the extraction of spectra from Wilson loops \cite{Burnier:2013fca}, and it needed the development of a new Bayesian reconstruction prescription \cite{Burnier:2013nla} before quantitatively robust results were obtained. Even with this new method the spectrum can be determined only in a certain range of frequencies, given by the energies resolved on the lattice, which makes a brute force evaluation of \eqref{Eq:PotSpec} impossible. A selection of reconstructed spectra for the two extreme cases of the lowest and highest investigated temperatures around $T_C$ are shown in fig.~\ref{Fig:WlineSpec}.

This further difficulty can also be overcome from a careful inspection of the spectral structure of the Wilson loop. Just as we did to arrive at eq.~\eqref{Eq:VRealTimeDef} let us assume that a potential picture is valid, i.e. the late time evolution of $W_\square(t,r)$ is dominated by a time independent function $\lim_{t\to\infty}\Phi(t,r)=V(r)$. As was proven in \cite{Burnier:2012az}, in this case the Wilson loop spectrum will contain a well defined lowest lying peak of skewed Lorentzian form embedded in a polynomial background
\begin{align}
&\rho\propto\frac{|{\rm Im} V(r)|{\rm cos}[{\rm Re}{\sigma_\infty}(r)]-({\rm Re}V(r)-\omega){\rm sin}[{\rm Re} {\sigma_\infty}(r)]}{ {\rm Im} V(r)^2+ ({\rm Re} V(r)-\omega)^2}\label{skewedrho}\\ \notag&+{c_0}(r)+{c_1}(r)({\rm Re} V(r)-\omega)+{c_2}(r)({\rm Re} V(r)-\omega)^2\ldots. 
\end{align}
The position and width of this peak are related to the real- and imaginary part of the potential respectively. The skewness and background contributions on the other hand are identified with remnants of the physics at scales above the soft-scale. 

We can now turn this relation around and use it as a non-perturbative criterion for the applicability of a potential description. If the Wilson loop spectrum does contain a well defined lowest lying peaked feature of skewed Lorentzian type, it will lead to a late-time evolution governed by a well defined potential.  As shown in the examples of reconstructed spectra in fig.~\ref{Fig:WlineSpec} we do indeed find such well pronounced peaked features and trough a fit establish that they of a skewed Lorentzian form\footnote{If the peaks were instead Gaussian $\rho_G(\omega)=c_0{\rm exp}[-(\omega-m)^2/\Gamma^2]$, inserting them into the defining formula \eqref{Eq:PotSpec} would lead to a divergent expression, since $\int\,d\omega\, \omega \rho_G(\omega) e^{-i\omega t} / \int \, d\omega\, \rho_G(\omega) e^{-i\omega t}   = m- i\frac{\Gamma^2}{2}t$ shows an unphysical linear increase in $\Im[V]$ over time \cite{Rothkopf:2011db}.}. This leads us to conclude that an in-medium potential description is warranted and the values of $V(r)$ can be extracted via the application of eq.\eqref{skewedrho}.

In practice, we look for the lowest lying peak in the spectra reconstructed from the lattice QCD observables, fit it around the full width at half maximum with the skewed Lorentzian form (\ref{skewedrho}) and read off the value of ${\rm Re}[V]$ and ${\rm Im}[V]$ encoded in it. This extraction strategy laid out above has been successfully tested using hard-thermal loop perturbation theory \cite{Burnier:2013fca}, where both the Wilson loop in Euclidean time, its spectrum and the corresponding potential are known. 

In this study we use the values of the heavy quark potential \cite{Burnier:2014ssa} extracted\footnote{Note that the observable we use here to extract the potential is not the Wilson loop but the Wilson line correlator in Coulomb gauge. The reason is that the latter does not contain cusp divergences that make evaluating the Wilson loop very costly on the lattice. We have checked that the result remains gauge invariant by relaxing the Coulomb condition and applying random gauge transformations.} from full QCD lattices generated by the HotQCD collaboration \cite{Bazavov:2011nk}, containing dynamical u,d, and s quarks. The medium quarks on these lattices with spatial extend $N_s=48$ are described by the asqtad action and are tuned to lie on a line of constant physics with a physical strange quark mass and light u,d quark masses close to their physical values  $m_s/m_{u,d}=20$ (For more details see \cite{Bazavov:2011nk}). In addition to the values of the potential around the phase transition at $T/T_C=0.68-1.66$, which had been extracted in a previous study \cite{Burnier:2014ssa}, we add here the values from two additional low temperature ensembles close to $T\sim0$, which will be used to calibrate the analytic expression for the temperature dependence of the potential in the next subsection. tab.~\ref{Tab:LatParm} summarizes the relevant simulation parameters for our ensembles and gives the temperatures in relation to the chiral crossover transition temperature on these lattices at $T_C=172.5$~MeV.

\begin{table}
\centering
\begin{tabularx}{15.5cm}{ |>{\centering}m{2cm} | >{\centering}m{1cm}| >{\centering}m{1cm}|| >{\centering}m{1cm} | >{\centering}m{1cm} | >{\centering}m{1cm} | >{\centering}m{1cm} | >{\centering}m{1cm} | >{\centering}m{1cm} | X | }
\hline
	$\beta$ \hspace{0.8cm} & 6.9 & 7.48 & 6.8 & 6.9 & 7 & 7.125 & 7.25 & 7.3 & 7.48 \\ \hline\hline
	$T$[MeV] & $\sim 0$ & $\sim 0$ & 148 & 164 & 182 & 205 & 232 & 243 & 286 \\ \hline
	$T/T_C$ & $\sim 0$ & $\sim 0$ & 0.86 & 0.95 & 1.06 & 1.19 & 1.34 & 1.41 & 1.66 \\ \hline
	a [fm] & 0.1 & 0.057 & 0.111 & 0.1 & 0.09 & 0.08 & 0.071 & 0.068 & 0.057\\ \hline
	$48^3\times N_{\tau}$ & 48 & 48 & 12 & 12 & 12 & 12 & 12 & 12 & 12\\ \hline
	$N_{\rm meas}$ &350 & 163 & 1295 & 1340 & 1015 & 1270 & 1220 & 1150 & 1130 \\ \hline
\end{tabularx}
\caption{Parameters of the isotropic HotQCD $48^3\times N_\tau$ lattices \cite{Bazavov:2011nk} with asqtad action ($m_l=m_s/20,T_c\approx172.5$MeV) used in this study.}\label{Tab:LatParm}
\end{table}

\begin{figure}
\centering
 \includegraphics[scale=0.35]{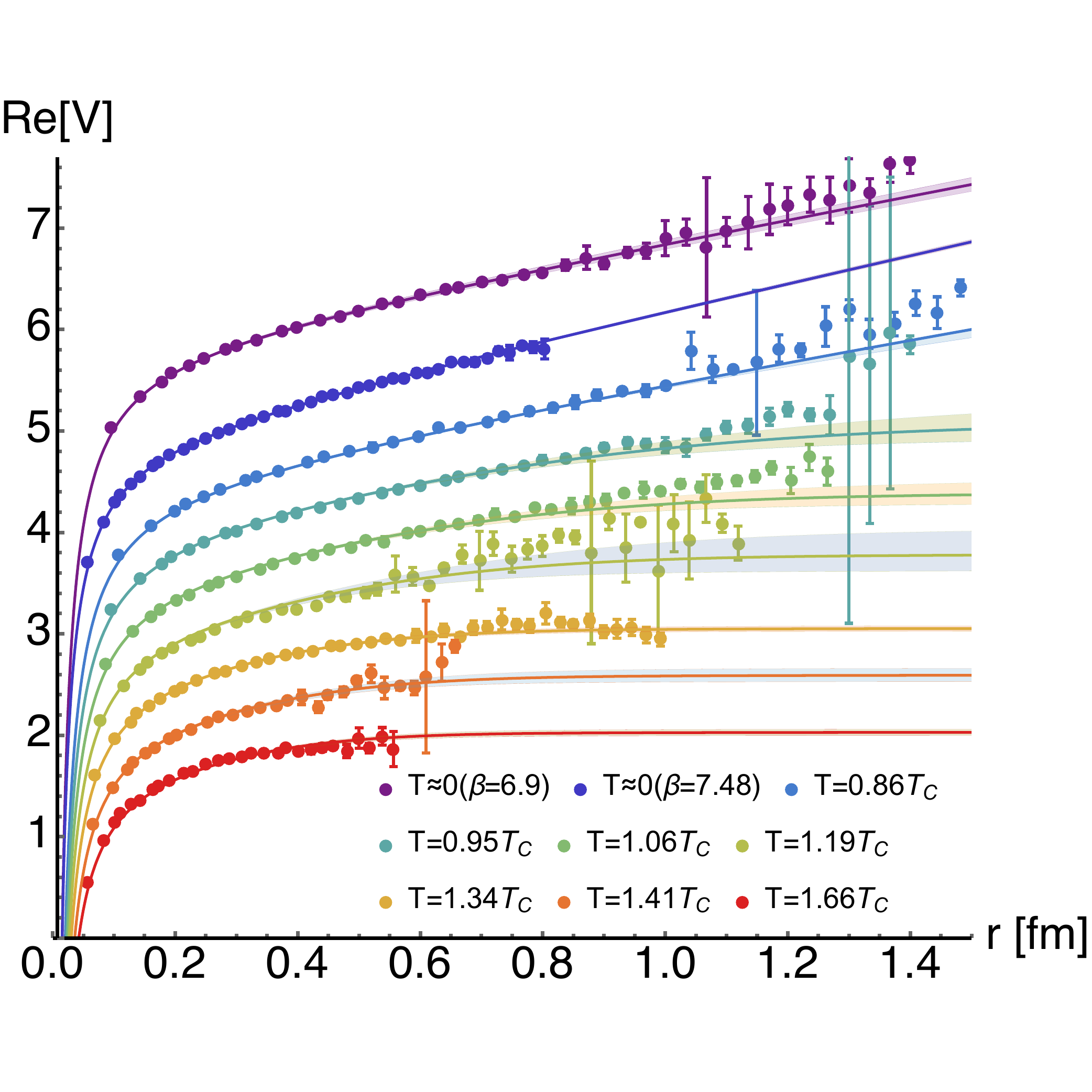}\hspace{0.2cm}
 \includegraphics[scale=0.35]{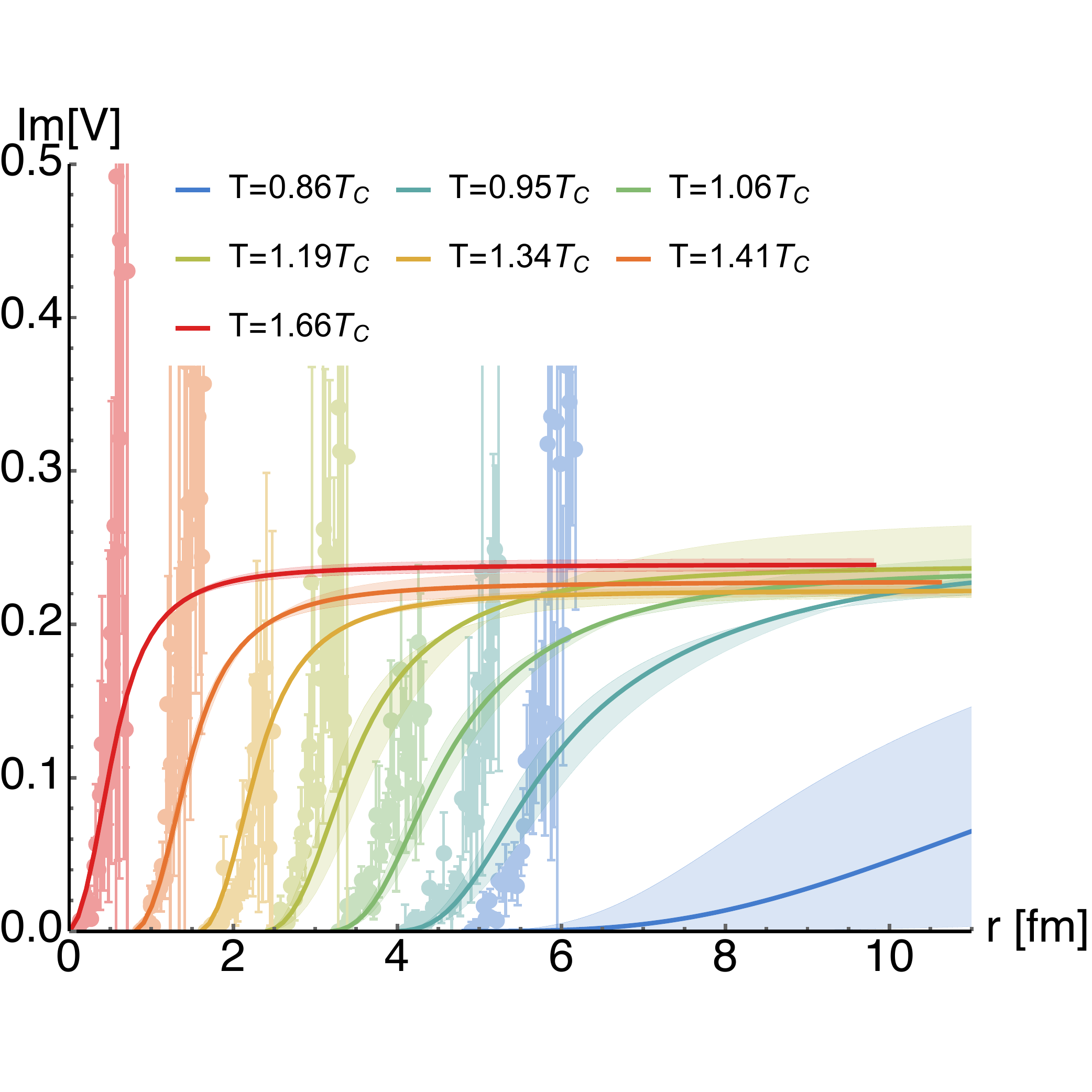}
 \caption{(left) Real-part of the in-medium heavy quark potential on $N_f=2+1$ asqtad lattices (points) with fits that establish the values of the Debye mass (solid line). Errorbands denote changes from varying the value of $m_D$ within its fit uncertainty. (right) prediction of the imaginary part of the potential (solid curves), together with the tentative values (light points) extracted from the asqtad lattices with $N_\tau=12$.}\label{Fig:ReImPot}
\end{figure}

Since the finite temperature lattices only contain $N_\tau=12$ lattice points in Euclidean time direction, we expect that while a robust determination of spectral peak positions is possible, the accuracy for spectral widths will be insufficient to make reliable statements about ${\rm Im}[V]$. The values obtained for the real-part are given by the colored points in the left panel of fig.~\ref{Fig:ReImPot}, the tentative values for ${\rm Im}[V]$ from the spectral widths are plotted as lightly colored points in the background of the right panel.

We will postpone the discussion of the temperature dependence of the potential to the end of the next subsection. 

\subsection{Gauss law parametrization of the potential}
\label{sec:GaussLaw}

In order to investigate heavy quarkonium spectra by solving a Schr\"odinger equation with the proper complex heavy-quark potential, we require an easily evaluable expression for both ${\rm Re}[V]$ and ${\rm Im}[V]$ at a given temperature. To this end we have recently proposed a field theoretically motivated functional form, that depends on a single temperature dependent parameter, the Debye screening mass \cite{Burnier:2015nsa}.

The idea behind this approach is as follows. The physics of heavy quarkonium at zero temperature can be described in a potential picture with two distinct characteristics, a Coulombic part at small distances and a linearly rising part at large distances\footnote{Note that by allowing the linear term to contribute down to the shortest distances we partially absorb the running of the strong coupling}. Hence if we wish to understand the in-medium modification of the inter-quark potential we need to understand how a test charge associated with either one of these two distinct potentials will react to the presence of medium charge carriers in its surroundings. To describe the effect of these charges we will use the hard-thermal loop medium permittivity, in essence making the ansatz of a test particle with a particular electric field configuration being immersed in a bath of weakly interacting quarks and gluons.

The basis for the derivation of the analytic expression for the complex potential lies in the generalized Gauss law derived in \cite{Dixit:1989vq}
\beq
\vec\nabla \left(\frac{\vE}{r^{a+1}}\right)=   4\pi \,q\, \delta(\vr) \label{Eq:GenGauss}.
\eeq
It is applicable to a point charge with (color) electric field $\vE=q r^{a-1} \rh$, which covers both the Coulombic $a=-1, q=\tilde\alpha_s, [\tilde\alpha_s]=1$, as well as string-like potential $a=1, q=\sigma, [\sigma]={\rm GeV}^2$ relevant for heavy quarkonium. In case of a Coulombic potential we can incorporate the in-medium effects in a straight forward manner by transforming \eqref{Eq:GenGauss} into Fourier space and multiplying the right hand side with a (possibly complex) permittivity, taken here from hard-thermal loop perturbation theory
\beq
\eps^{-1}(\vp,m_D)=\frac{p^2}{p^2+m_D^2}-i\pi T \frac{p\, m_D^2}{(p^2+m_D^2)^2}.\label{Eq:HTLperm}
\eeq
Transforming back to coordinate space yields
\beq
-\nabla^2 V_C(r)+m_D^2V_C(r)=\tilde\alpha_s \Big(  4 \pi  \delta(\vr)-  iT m_D^2 \varphi(m_Dr)\Big),\label{Eq:Coulomb(r)withIm}
\eeq
with
\beq
\varphi(x)=2\int_0^\infty dp  \frac{\sin(p x)}{p x}\frac{p}{p^2+1},
\eeq
which constitutes an integro-differential equation with linear-response character for the in-medium modified Coulomb potential. The strength of the medium effects is clearly governed by the parameter $m_D$, which we have proposed as a non-perturbative definition of the Debye mass in ref.~\cite{Burnier:2015nsa}. Solving \eqref{Eq:Coulomb(r)withIm} with the appropriate boundary conditions $\left.{\rm Re}V_C(r)\right|_{r=\infty}=0$, $\left.{\rm Im}V_C(r)\right|_{r=0}=0$ and $\left.\partial_r {\rm Im}V_C(r)\right|_{r=\infty}=0$ reproduces the perturbative potential of Laine et. al \cite{Laine:2007qy} given in eq.\eqref{Eq:VHTL}. 

In the case of a string-like potential, transforming eq.\eqref{Eq:GenGauss} into Fourier space is impractical. Instead we rely on the assumption that the in-medium charge distribution found in the Coulomb case is the same as for the linearly rising potential i.e. a property of the medium and not of the test charge. As was shown in \cite{Burnier:2015nsa} the strength of the medium effects in the resulting linear-response type equation is governed not by the Debye mass alone but instead by the parameter $\mu^4=m_D^2\frac{\sigma}{\tilde\alpha_s}$
\beq
-\frac{1}{r^{2}}\frac{d^2 V_s(r)}{dr^2}+ \mu^4 V_s(r)=\sigma \Big( 4 \pi \delta(\vr)- i T m_D^2 \varphi(m_D r)\Big).\label{Eq:String(r)withIm}
\eeq
Solving for the real part of the medium modified string potential leads to an analytic expression in terms of parabolic cylinder functions $D_\nu(x)$
\begin{align}
{\rm Re}V_s(r)&=-\frac{\Gamma[\frac{1}{4}]}{2^{\frac{3}{4}}\sqrt{\pi}}\frac{\sigma}{\mu} D_{-\frac{1}{2}}\big(\sqrt{2}\mu r\big)+ \frac{\Gamma[\frac{1}{4}]}{2\Gamma[\frac{3}{4}]} \frac{\sigma}{\mu}.
\end{align}
The imaginary part on the other hand  can be written in a closed form using the Wronskian method, which yields
\begin{eqnarray}
\Im V_s(r)&=&-i\frac{\sigma m_D^2 T}{\mu^4} \psi(\mu r)=-i\tilde\alpha_s T \psi(\mu r)\label{Eq:ImVSGenGauss},
\end{eqnarray}
with
\begin{eqnarray} 
\notag\psi(x)&=&D_{-1/2}(\sqrt{2}x)\int_0^x dy\, {\rm Re}D_{-1/2}(i\sqrt{2}y)y^2 \varphi(ym_D/\mu)\\\notag&&+{\rm Re}D_{-1/2}(i\sqrt{2}x)\int_x^\infty dy\, D_{-1/2}(\sqrt{2}y)y^2 \varphi(ym_D/\mu)\\&&-D_{-1/2}(0)\int_0^\infty dy\, D_{-1/2}(\sqrt{2}y)y^2 \varphi(ym_D/\mu).
\end{eqnarray}

Note that even though we have used a weak coupling ansatz for the permittivity, which is only appropriate at high temperatures, its insertion into on the Gauss law has produced an expression for the potential with linear-response character. In turn it seems to smoothly connect to zero temperature, as the value of the Debye mass can in principle be set to zero. In particular the real part of the Gauss law derived in-medium potential reduces to the $T=0$ Cornell-potential at $m_D=0$. This bodes well for applying the derived expression to fitting the lattice QCD extracted potential values even at temperatures below the formal range of applicability of weak-coupling methods. 

\subsection{Determination of $m_D$ from the lattice potential}
\label{sec23}

For a consistent determination of the single temperature dependent parameter $m_D$, we assume that neither the strong coupling nor the string tension depend on the medium temperature, i.e. all modification emerges from the surrounding light quarks and gluons. Thus we will need to fix the values of $\tilde\alpha_s$, $\sigma$, as well as the arbitrary scale dependent constant shift of ${\rm Re}[V]$ at $T=0$. Since lattice QCD simulations are performed on finite size lattices at a finite lattice spacing, they necessarily operate at a low but finite temperature. In our case we will hence use the newly added data from the two ensembles close to zero temperature at $\beta_1=6.9$ and $\beta_2=7.48$. We find that the values of the vacuum parameters vary slightly between the two ensembles. Hence a linear interpolation is used at intermediate lattice spacings.\footnote{Note that the differences between the values of $\sigma$ at different $\beta$ values might be related to an insufficiently precise setting of the scale for the asqtad lattices in \cite{Bazavov:2011nk}} As can be seen from the dark and light blue curves at the top of the right panel of fig.~\ref{Fig:ReImPot} the lattice values for ${\rm Re}[V]$ at low temperature indeed show a well pronounced Coulombic and linear behavior that is excellently reproduced by the fit parameters given in tab.~\ref{Tab:VacParm}. I.e. neither the running of the coupling at small distances nor logarithmic corrections at large distances are significant for the regime investigated here.

\begin{table}[h!]
\centering
 \begin{tabular}{c|c|c}
  ~ &$\beta$=6.9& $\beta$=7.48\\ \hline
  $\tilde{\alpha}_s$ & $0.456\pm0.027$ & $0.385\pm0.006$ \\
  $\sqrt{\sigma}$ [GeV] & $0.470\pm0.01$ & $0.515\pm0.003$\\
  c [GeV]& $1.760\pm0.034$ & $2.648\pm0.009$
 \end{tabular}
 \caption{Values for the vacuum potential parameters from the low temperature lattice QCD ensembles}\label{Tab:VacParm}
\end{table}

\begin{figure}
\centering
 \includegraphics[scale=0.55]{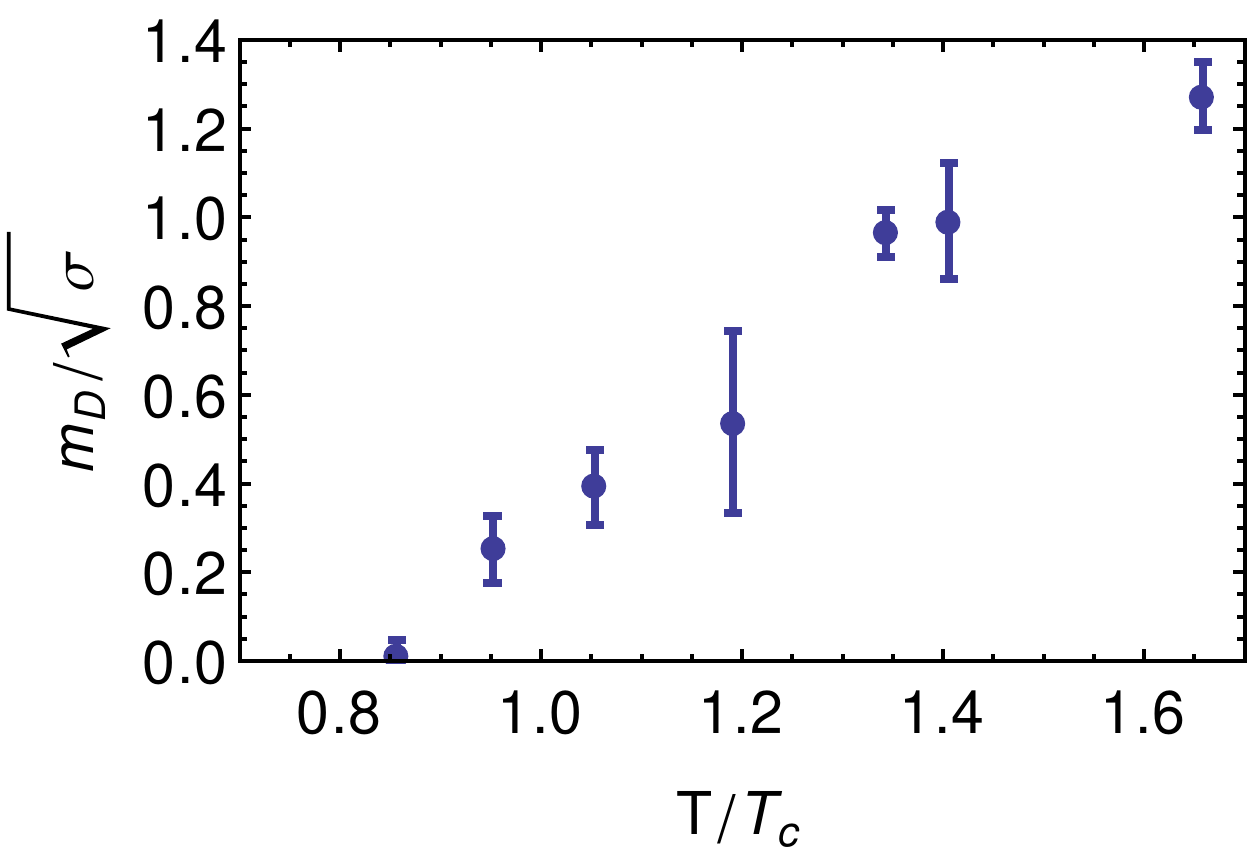}\hspace{0.2cm}
 \includegraphics[scale=0.55]{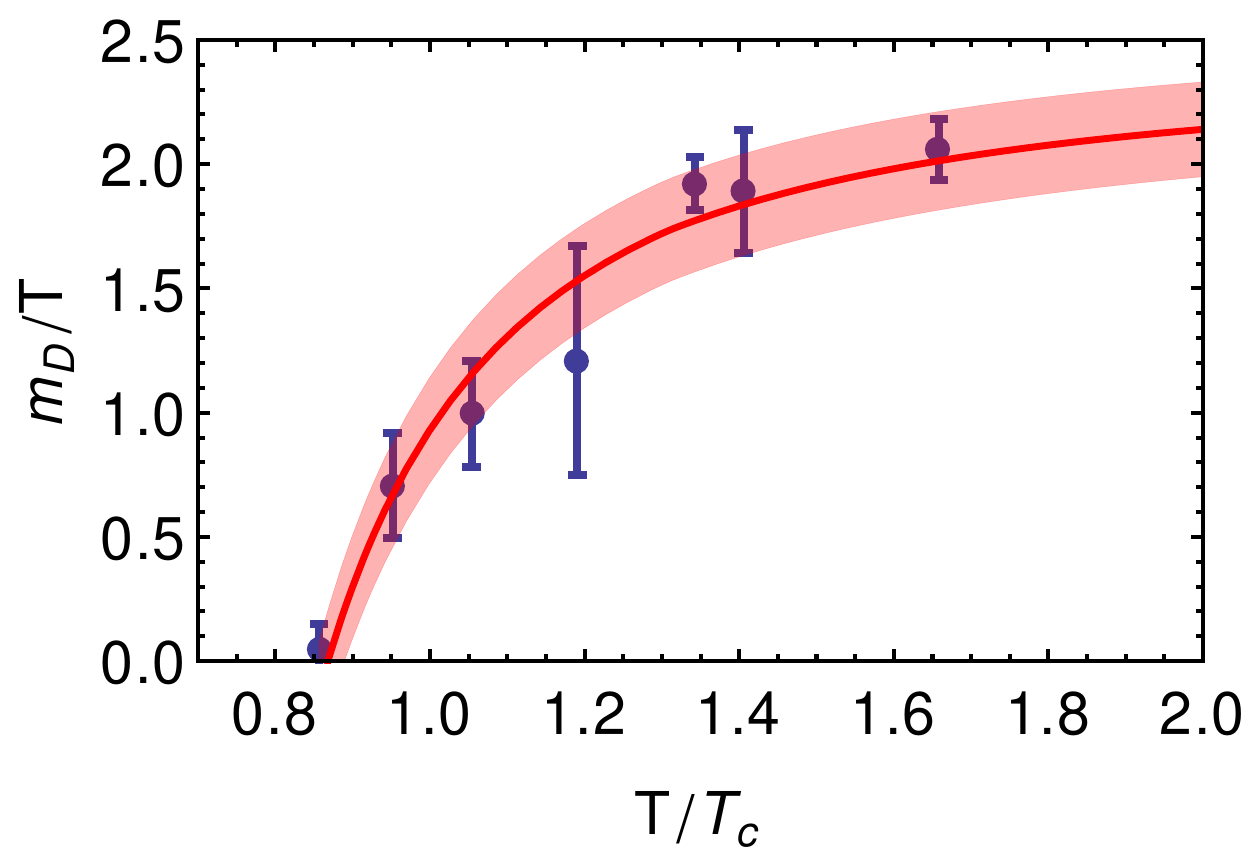}
 \caption{(left, blue points) The normalized Debye mass ($m_D/\sqrt{\sigma}$) from a generalized Gauss-law fit to the real-part of the in-medium heavy quark potential on asqtad lattices, which we propose as input for phenomenological studies. (right), Blue points: temperature dependence of the Debye mass ($m_D/T$) and in red a NLO HTL based fit of $m_D$}\label{Fig:mDFits}
\end{figure}

With the vacuum values set and since the explicit expression for both ${\rm Re}[V]$ and ${\rm Im}[V]$ derived above only depends on a single parameter, we continue by determining $m_D$ from a fit to the real part of the lattice QCD extracted values alone. The resulting curves are given as solid lines in the right panel of fig.~\ref{Fig:ReImPot}, the corresponding Debye masses are collected in tab.~\ref{Tab:DebMass} and plotted in fig.~\ref{Fig:mDFits}. 

\begin{table}[h]
\centering
\begin{tabularx}{15.5cm}{ |>{\centering}m{1cm} | >{\centering}m{1.6cm}| >{\centering}m{1.6cm} | >{\centering}m{1.6cm} | >{\centering}m{1.6cm} | >{\centering}m{1.6cm} | >{\centering}m{1.6cm} | X | }
\hline
	$\beta$ \hspace{0.8cm} & 6.8 & 6.9 & 7 & 7.125 & 7.25 & 7.3 & 7.48 \\ \hline\hline
	$T/T_C$ & 0.86 & 0.95 & 1.06 & 1.19 & 1.34 & 1.41 & 1.66 \\ \hline
	$\frac{m_D}{\sqrt{\sigma}}$ & 0.01(3) & 0.25(8) & 0.39(8) & 0.53(21) & 0.96(5) & 0.99(13) & 1.27(8) \\ \hline
	$\frac{m_D}{T}$ & 0.04(10) & 0.72(22) & 1.03(22)&1.28(49)&2.07(11)&2.05(27)&2.29(14)  \\ \hline
\end{tabularx}
\caption{Debye masses extracted from the isotropic HotQCD $48^3\times12$ lattices with asqtad action. For use in phenomenology, a continuum corrected $m_D$ may be obtained from the ratio $\mD/\sqrt{\sigma(\beta)}$ shown here, through a multiplication with the continuum value of $\sigma$.}\label{Tab:DebMass}
\end{table}

We find that tuning $m_D$ allows us to indeed reproduce both the qualitative and quantitative behavior of ${\rm Re}[V]$ without problem, if the interpolated $T=0$ constants are used. The strength of the generalized Gauss law Ansatz is that it now allows us to predict the values of the imaginary part of the potential, which was deemed unreliable in a previous study due to the fact that on the finite temperature lattices the Wilson lines were available at only twelve points in Euclidean direction. We are confident in the predictive capabilities of the approach, as it has been able to successfully reproduce ${\rm Im}[V]$ in a similar study in quenched QCD \cite{Burnier:2015nsa}.

When plotting ${\rm Im}[V]$, as implied by $m_D(T)$ via eq.~\eqref{Eq:Coulomb(r)withIm} and eq.\eqref{Eq:ImVSGenGauss} as solid lines in the right panel of fig.~\ref{Fig:ReImPot} the tentative data (light colored points) shows a surprisingly good agreement with these values at small separation distances $r<0.75$fm down to $T=0.95T_C$. Only at  $T=0.86T_C$ we find large differences between the extracted and predicted values, which is probably due to the impossibility to resolve the tiny width of the spectral peaks at low temperatures using only $N_\tau=12$ data points. Since heavy quarkonium phenomenology in the quark gluon plasma requires knowledge about the potential only until the freezeout boundary is reached, i.e.  slightly below the deconfinement transition, the apparent range of applicability of the fit seems to suffice.

\subsection{Continuum correction for the potential and $m_D$}\label{pot_cont}

Several preparations are still in order, since for a meaningful phenomenological investigation, we need to use the continuum values for all parameters entering the potential. In the absence of a true continuum extrapolation of our lattice results we resort to the following strategy. 

We fit the vacuum potential parameters, $\tilde{\alpha}_s$, $\sigma$ and the constant term $c$ by comparing the energy levels of the corresponding Cornell-potential Hamiltonian to the experimentally measured masses of the bottomonium system, where we expect that finite mass correction are insignificant. More importantly for bottom quarks there exists a well controlled procedure to define their physical mass. A mass often deployed in pNRQCD effective theory calculations is the pole mass \cite{Melnikov:2000qh,Marquard:2007uj,Gray:1990yh} (for bottom $m_b^{\rm pole}=4.93$GeV) different from the usually quoted $\overline{MS}$ mass $(m^{\overline{MS}}_b(m^{\overline{MS}}_b)=4.18)$. Its larger value reflects the fact that the quark mass should be evaluated at the soft scale, relevant for the physics of the bound state and not at the hard scale. A renormalization group flow towards lower energies is hence required, leading to the larger $m_b^{\rm pole}$. More precisely, because of the so called renormalon ambiguity in the perturbative evaluation of the Wilson coefficients of the effective theory, one is lead to use the renormalon subtracted scheme \cite{Pineda:2001zq}. I.e. one subtracts the renormalon ambiguity in the pole mass definition and reshuffles it into the constant part of the potential, which suffers from the same ambiguity. 
In this way, the ambiguities eventually cancel and we are lead to a quantitatively robust definition of the potential. In general also here one obtains larger values than those resulting from the $\overline{MS}$ scheme, such as the established
\beq
m_b^{\rm RS'}=4.882\pm0.041 \,{\rm GeV}
\eeq
from ref.~\cite{Pineda:2001zq}, which has been deployed e.g. in \cite{Brambilla:2004jw}. The errors are estimated to be as large as the highest calculated contribution of order ${\cal O}(\alpha_S^4)$.

This choice of mass allows us to reproduce the PDG values for the four S-wave states $\Upsilon(1S)-\Upsilon(4S)$, the averaged massed of the three known P-wave triplets $\chi_b(1P)-\chi_b(3P)$ as well as the lowest D-wave state $\Upsilon(1D)$ to at least three significant digits if the remaining vacuum parameters are set to the following values
\beq
c=-0.1767\pm 0.0210~{\rm GeV}, \quad \tilde\alpha_s=0.5043\pm 0.0298,\quad \sqrt{\sigma}=0.415\pm0.015~{\rm GeV}.\label{T0const}
\eeq
These continuum values are very close to the ones used in conventional quarkonium spectroscopy \cite{Quigg:1979vr,Eichten:1995ch} and are also compatible with our lattice data. Note that the effects of string breaking were introduced by hand, flattening off the potential at $r_{sb}=1.25$fm \cite{Bali:2005fu}. A naive complex scaling analysis confirms that the four lowest S-wave states of the modified Cornell potential Hamiltonian indeed lie below the continuum threshold.

Lattice QCD simulations are carried out in a finite box with a finite lattice spacing, hence discretization effects have to be taken into account. The former is related to the fact that the quarks do not take on their physical masses exactly and thus the chiral crossover temperature on our lattices ($T_C^{\rm lat}=172.5$MeV) is higher than the continuum value of $T_C^{\rm cont}=155\pm9$MeV. In addition the change in lattice spacing leads to slightly different values of the vacuum parameters of the potential in our case, as discussed in sec.~\ref{sec23}. Hence both effects should be corrected for. Ultimately, we will use the constants of eq.(\ref{T0const}) together with the continuum corrected Debye mass 
\beq
m_D^{\rm phys}(t=T/T_C^{\rm phys})=\frac{m_D\left(t\right)}{\sqrt{\sigma(\beta)}}\sqrt{\sigma^{\rm cont}}\label{mDcont}
\eeq
where $\frac{m_D}{\sqrt{\sigma(\beta)}}$ is given in tab.~\ref{Tab:DebMass} and $\sigma^{\rm cont}$ in (\ref{T0const}).

Let us take a closer look at the continuum corrected Debye masses and how they compare to perturbative estimates. In the work of ref.~\cite{Arnold:1995bh} it has been established that the Debye mass can be computed perturbatively only up to the leading order together with the logarithmic correction at next to leading order (NLO). The presence of a magnetic sector in QCD leads to the appearance of truly non-perturbative contributions to $m_D$ at NLO, parametrized in the following by the two terms containing the constants $\kappa_1, \kappa_2$, which need to be determined from numerical simulations
\begin{eqnarray}
m_D&=&T g(\mu)\sqrt{\frac{N_c}{3}+\frac{N_f}{6}} \notag
	+\frac{N_c T g(\mu )^2}{4 \pi } \log
   \left(\frac{\sqrt{\frac{N_c}{3}+\frac{N_f}{6}}}{g(\mu )}\right)
   \\&&+\kappa_1 \,T g(\mu )^2+\kappa_2\, T g(\mu )^3. \label{Eq:mD}
\end{eqnarray}
To compare the temperature dependence of the Debye mass, we fix the non-perturbative constants $\kappa_1,\kappa_2$ while keeping $\mu=2\pi T$ constant. For the running of the coupling $g(\mu)$ we utilize the four loop result of ref.~\cite{Vermaseren:1997fq} setting $\Lambda_{QCD} = 0.2145$GeV, appropriate for starting the renormalization group flow from a scale where $N_f=5$ flavors are active. 

In the left panel of fig.~\ref{Fig:mDFits} we show the values of $m_D/T$ (blue points) together with the fit according to eq.~\eqref{Eq:mD}. Even though the perturbative part of the formula for $m_D$ is applicable, if at all, at the highest temperatures investigated here, we find that the fit manages to pass through all values of $m_D$ even around the phase boundary. The non-perturbative contributions are non-negligible with $\kappa_1= 0.84\pm0.10$ and $\kappa_2=-0.40\pm0.03$. Nevertheless the perturbation theory motivated fit (by chance) reproduces $m_D$ down to temperatures, slightly below $T_c$ and may hence be used to define a phenomenological, lattice QCD validated, temperature dependence of the Debye screening mass.

\section{Quarkonium spectra}
\label{section3}

In sec.~\ref{section2} we managed to capture the functional form of both the real- and imaginary-part of the static potential by fitting a single temperature dependent parameter $m_D$, the Debye mass, to ${\rm Re}[V]$ extracted on the lattice. We now take the next step and compute from it the in-medium spectral functions of the vector channel bottomonium and charmonium S-wave states at finite mass by solving an appropriate Schr\"odinger equation. These spectra correspond to resting states with absolute momenta $\mathbf{p}=0$, similar to those usually investigated in direct lattice QCD or lattice NRQCD studies. 

The static heavy quark potential is a universal quantity, in the sense that it denotes the lowest order contribution in the non-relativistic expansion for both bottomonium and charmonium physics. Therefore we expect that the vacuum parameters fitted in the bottomonium case do remain the same for the lighter flavor. An important difference between the two cases is that the binding energy of the charmonium ground state is close to $\Lambda_{\rm QCD}$. This makes it impossible to set up a perturbative renormalization scheme for the charm mass,  similar to the one we used for Bottom. Hence instead of calculating the renormalon subtracted mass, we use the charm mass as fit parameter and tune it to reproduce the masses of the stable S-wave states ($J/\Psi,\Psi'$) and the averaged two P-wave triplets $\chi_c(1P),\chi_c(2P)$. The resulting best fit value reads 
\beq
m_c^{\rm PDG\,fit}=1.472\, {\rm GeV}.\label{Eq:CharmMass}
\eeq 
Since finite mass effects, in particular radiative corrections can become relevant for charmonium, we expect the agreement between the static potential based masses and the experimental $T=0$ spectrum to be worse than for bottomonium. Indeed for the S-wave states and the 1P triplet only agreement up to the second digit is found when these higher order effects are neglected.

The vacuum parameters determined in sec.~\ref{pot_cont} and \eqref{Eq:CharmMass} constitute the basis from which we embark on the finite temperature study, where all medium effects on the static potential are summarized in the temperature dependence of the Debye mass parameter determined in sec.~\ref{sec23}. We use the continuum string tension of the bottomonium fit to convert the values of $m_D/\sqrt{\sigma}$ of tab.~\ref{Tab:DebMass}, i.e. the right panel of fig.~\ref{Fig:mDFits} for use in the continuum Schr\"odinger equation. Note that we have set up the potential parametrization in sec.~\ref{section2} such that changing the Debye mass does not affect the overall constant in ${\rm Re}[V]$.  I.e. at small enough separation distances, where temperature effects are irrelevant, the values of ${\rm Re}[V]$ all agree independently of $m_D$, which is expected of a correctly renormalized potential and a necessary requirement for a meaningful interpretation of the quarkonium bound state physics at finite $T$. tab.~\ref{Tab:BottomParm} and tab.~ \ref{Tab:CharmParm} summarize the vacuum properties of several bottomonium and charmonium states that arise from the $T=0$ potential parameters.

\begin{table}
\centering
\begin{tabularx}{16.5cm}{ |>{\centering}m{2.7cm} | >{\centering}m{1.5cm}| >{\centering}m{1.5cm}| >{\centering}m{1.5cm} | >{\centering}m{1.5cm} || >{\centering}m{1.5cm} | >{\centering}m{1.5cm} | X | }
\hline
 states    & $\Upsilon(1S)$ & $\Upsilon(2S)$ & $\Upsilon(3S)$ & $\Upsilon(4S)$ & $\chi_b(1P)$ & $\chi_b(2P)$ & $\chi_b(3P)$  \\ \hline\hline
 $m$ [GeV] & 9.4603 & 10.020 & 10.353 & 10.597 & 9.92597 & 10.269 & 10.538 \\ \hline
 $m^{\rm \scriptscriptstyle PDG}$ [GeV]& 9.4603 & 10.023 & 10.355& 10.579 & 9.88814 & 10.252 & 10.534\\ \hline
 $\langle r \rangle$  [GeV$^{-1}$]&  1.489 & 2.985 & 4.385 & 10.17& 2.435& 3.898& 5.586 \\ \hline
 $\langle r \rangle$ [fm] & 0.2934 & 0.5881 & 0.8639 & 2.004 & 0.4797& 0.7679& 1.100\\ \hline
 $\bar{m}_{\rm B\bar{B}}^{\rm \scriptscriptstyle PDG}-m$[GeV]& 1.1 & 0.539 & 0.206 & -0.038 & 0.633& 0.29 & 0.02\\ \hline
\end{tabularx}
\caption{The masses, mean radii and distances to the $B\bar{B}$ threshold for bottomonium S-wave and P-wave states at T=0.}\label{Tab:BottomParm}
\end{table}

\begin{table}
\centering
\begin{tabularx}{11.5cm}{ |>{\centering}m{2.7cm}|>{\centering}m{2.2cm} | >{\centering}m{1.5cm} || >{\centering}m{1.5cm} | X | }
\hline
 states    & $J/\Psi(1S)$ & $\Psi'(2S)$ & $\chi_c(1P)$ & $\chi_c(2P)$   \\ \hline\hline
 $m$ [GeV] & 3.0969 & 3.6717 & 3.5089 & 3.7918 \\ \hline
 $m^{\rm \scriptscriptstyle PDG}$ [GeV]& 3.0969 & 3.6861 & 3.4939 & 3.9228\\ \hline
 $\langle r \rangle$ [GeV$^{-1}$]&  2.861 & 5.839 & 4.136 & 25.42 \\ \hline
 $\langle r \rangle$ [fm] & 0.5635 & 1.150 & 0.814824 & 5.00813\\ \hline
 $\bar{m}_{\rm D\bar{D}}^{\rm \scriptscriptstyle PDG}-m$[GeV]& 0.639 & 0.064 & 0.227 & -0.056\\ \hline
\end{tabularx}
\caption{The masses, mean radii and distances to the $D\bar{D}$ threshold for charmonium S-wave and P-wave states at T=0.}\label{Tab:CharmParm}
\end{table}

\FloatBarrier

\subsection{Spectral functions from the Schr\"odinger equation}
\label{sec:SpecFunc}

All ingredients have been assembled for computing the vector channel spectral functions from the time evolution of the corresponding correlation function $D^>(t,\mathbf{r},\mathbf{r}')$ governed by the complex in-medium potential\footnote{Note again that it is not the Schr\"odinger equation for the quarkonium wavefunction but for the forward correlator that we are solving here.}. In the following we deploy the Fourier space method developed in ref.~\cite{Burnier:2007qm}, which solves
\begin{align}
\Big[ \tilde{H} - i|\Im V(r)| \Big] D^>(t,\mathbf{r},\mathbf{r}')=i\partial_t D^>(t,\mathbf{r},\mathbf{r}'),\quad t>0\\
\Big[ \tilde{H} + i|\Im V(r)| \Big] D^>(t,\mathbf{r},\mathbf{r}')=i\partial_t D^>(t,\mathbf{r},\mathbf{r}'),\quad t<0
\end{align}
with
\begin{align}
 \tilde{H}= 2m_Q -\frac{\nabla^2}{2m_Q} +{\rm Re}[V](r)+\frac{l(l+1)}{m_Qr^2} 
\end{align}
and the starting condition
\beq
D^>(0,\mathbf{r},\mathbf{r}')=-6N_c \delta^{(3)}(r-r').
\eeq
For the correlator in frequency space we Fourier transform
\begin{align}
\tilde{D}(\omega,\mathbf{r},\mathbf{r}')\equiv \int_{-\infty}^{\infty}dts^{i\omega t} D^>(t,\mathbf{r},\mathbf{r}'),
\end{align}
from which the vector channel spectrum is obtained by taking the limit
\begin{align}
 \rho^{\rm V}(\omega)=\lim_{\mathbf{r},\mathbf{r}'\to \mathbf{0}}\frac{1}{2} \tilde{D}(\omega,\mathbf{r},\mathbf{r}').
\end{align}
Taking the point splitting to zero requires some effort in the practical implementation as discussed in appendix A of ref.~\cite{Burnier:2007qm}.

In fig.~\ref{Fig:ChrmBotRhoOvervwS} we give overview plots of the computed in-medium S-wave spectral functions for bottomonium and charmonium at several temperatures around the transition temperature.  Note that the widths seen here are actual physical widths, related to finite temperature effects, i.e. all bound states reduce to delta peak structures in the $T=0$ limit. By construction, the $T=0$ peaks coincide with the experimental values for the bottomonium states up to 3 digits and fit the two charmonium states below threshold up to few percents \footnote{For $T=0$, we actually add a small imaginary part in the potential to avoid having exact delta-functions in the spectrum, that would be impossible to visualize graphically.}.

\begin{figure}
\centering
 \includegraphics[scale=0.75]{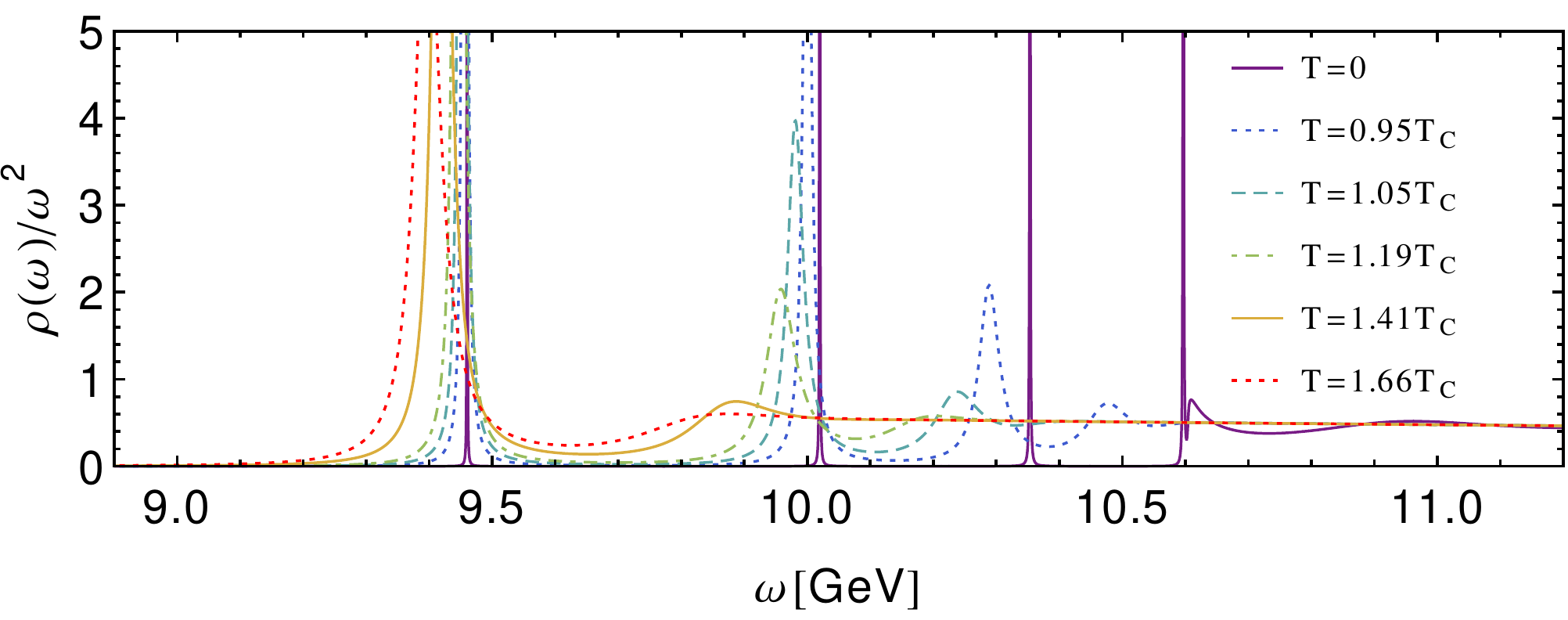}\hspace{0.2cm}
 \includegraphics[scale=0.75]{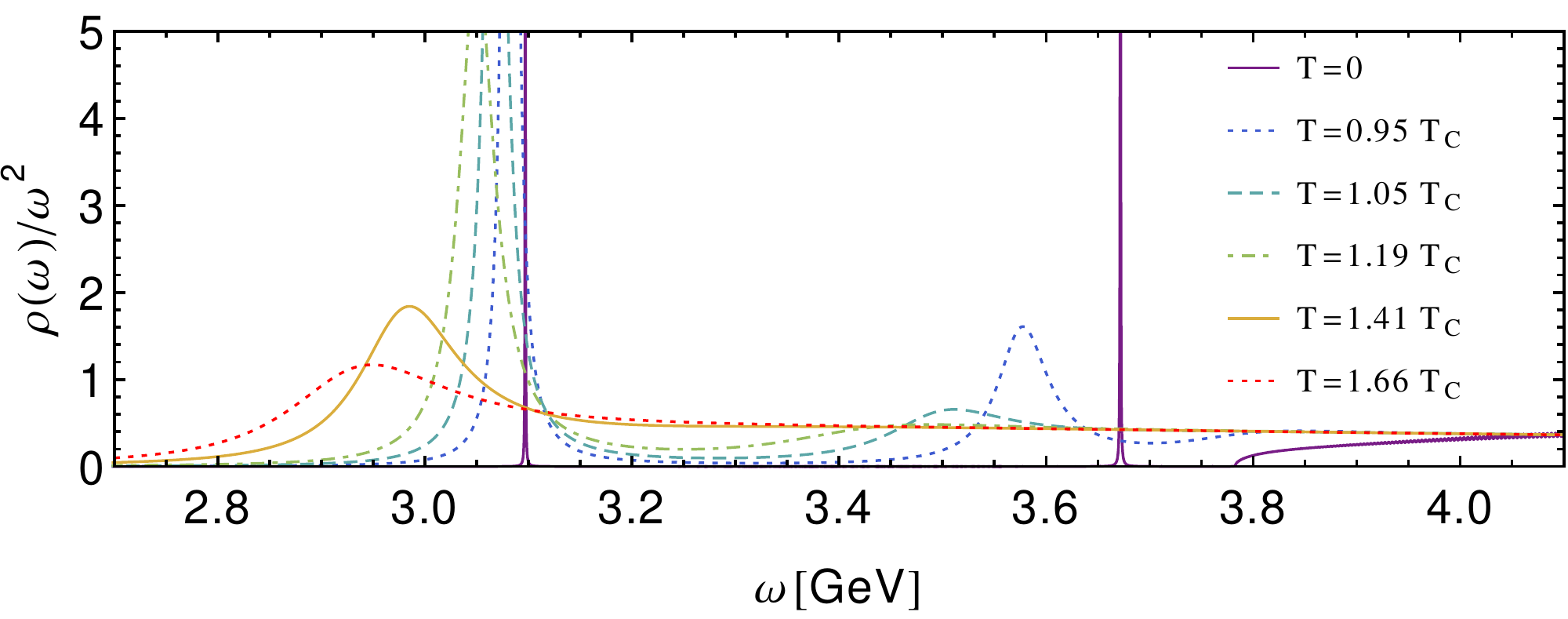}
 \caption{(top panel) S-wave bottomonium in-medium spectral function based on a proper complex potential as defined in sec.~\ref{pot_cont}. In this adiabatic scenario we find clear indications for sequential melting. $\Upsilon(4S)$ is almost gone already at $0.94T_C$, $\Upsilon(3S)$ disappears rapidly close to $T_C$ and $\Upsilon(2S)$ starts to be washed out at $1.19 T_C$.  Only $\Upsilon(1S)$ remains discernible at our highest lattice temperature $1.66T_C$.  (bottom panel) The charmonium in-medium spectral function based on the same complex potential as for bottomonium. In the absence of a well defined renormalon subtracted charm mass, its value is fitted to reproduce the known charmonium bound states at $T=0$ giving $m_c=1.472$GeV. Also charmonium shows a sequential melting pattern, as we find that $\Psi(2S)$ disappear around $T_C$ while $J/\Psi$ close to $1.41T_C$.
 }\label{Fig:ChrmBotRhoOvervwS}
\end{figure}

\subsection{Properties of the in-medium spectral functions}

As expected from the difference in constituent quark mass, the bottomonium states react less severely to the medium at a given temperature compared to charmonium. In general we can observe in this adiabatic setting that the very narrow peaks at low temperature are affected in a hierarchical manner, the highest lying states, which are most loosely bound, begin to broaden and shift first, followed sequentially by the lower lying states. The combination of the effects of screening $\Re[V]$ and scattering $\Im[V]$ encoded in the potential lead to characteristic common changes for both quark flavors. Not only do the bound states broaden but also their mass shifts to lower values. This behavior is found not only for the ground state but for all the higher lying states before they eventually dissolve and become part of the continuum. 

To be more quantitative, if a narrow resonance pole lies close to the real frequency axis its spectrum can be described by a simple Breit-Wigner (BW). On the other hand for states that are close to melting, it is necessary to disentangle the remnant bound state signal from the continuum background. For such broad features a skewed Breit-Wigner needs to be considered  \cite{Taylor}, which reads
\beq
\rho(\omega\approx E)=C \frac{(\Gamma/2)^2}{(\Gamma/2)^2+(\omega-E)^2}+2
\delta \frac{(\omega-E)\Gamma/2}{(\Gamma/2)^2
+(\omega-E)^2}+\mathcal{O}(\delta^2),\label{sBW}
\eeq
where $E$ denotes the energy of the resonance, $\Gamma$ its width and
$\delta$ the phase shift.

\begin{figure}
\centering
 \includegraphics[scale=0.585]{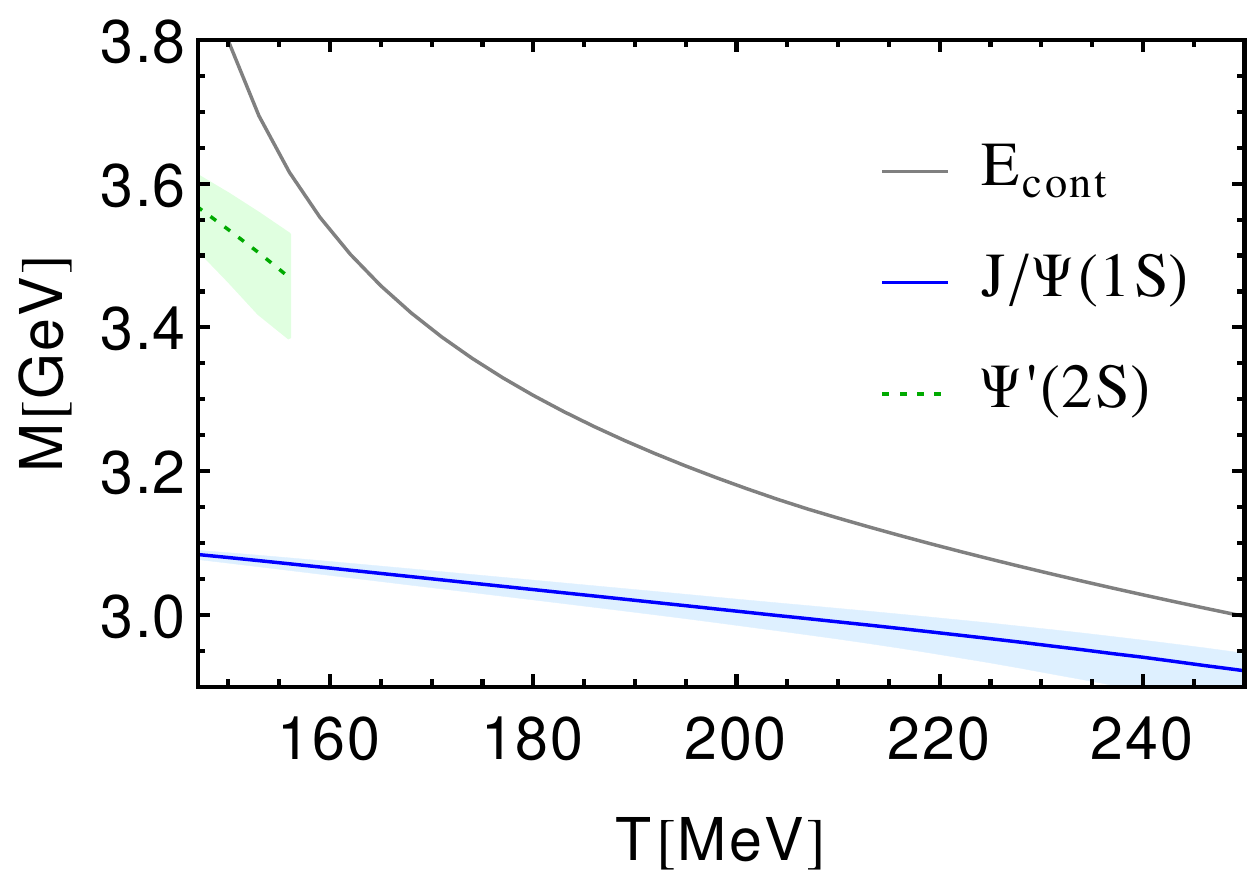}\hspace{0.2cm}
 \includegraphics[scale=0.6]{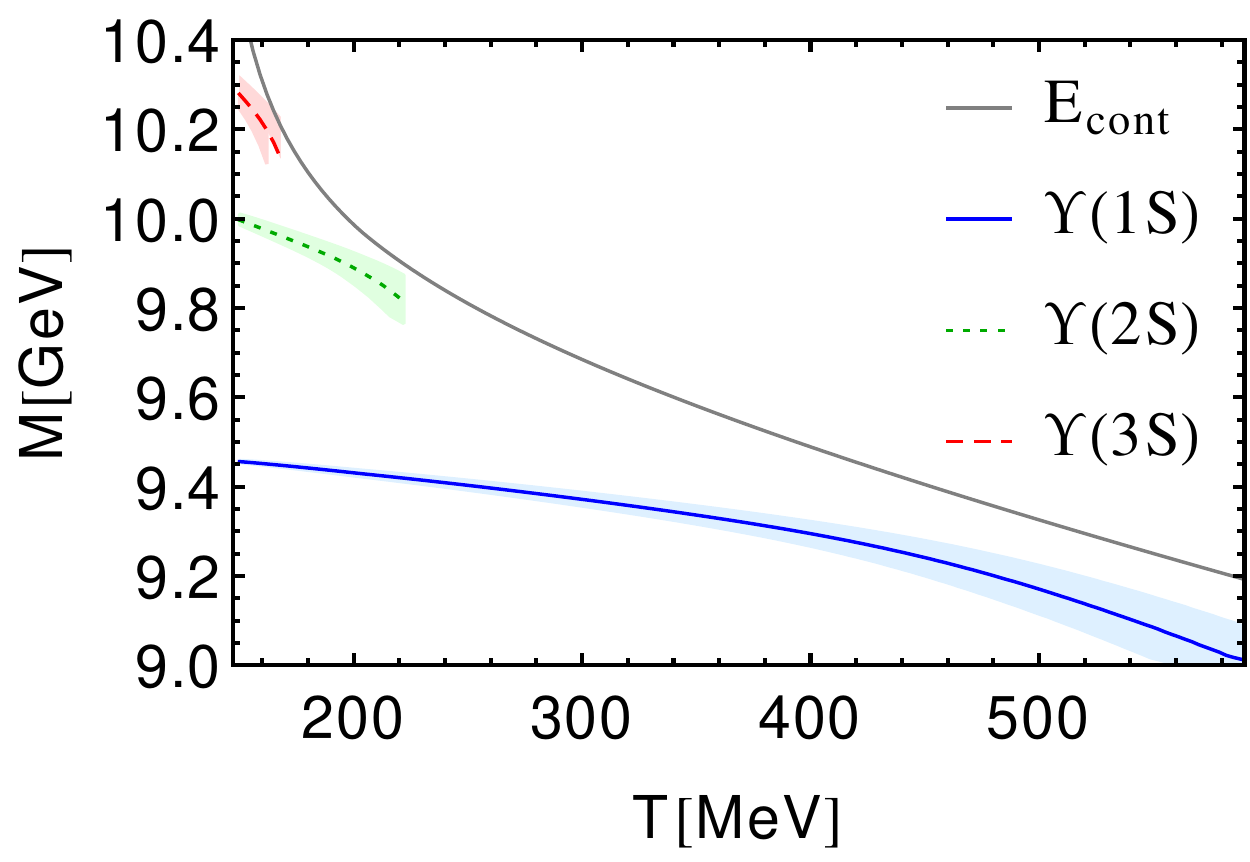}
 \caption{Mass of the charmonium (left) and bottomonium (right) bound states from the position of their in-medium spectral peaks. The peak position decreases monotonously with temperature until the bound state disappears. The continuum threshold energy $E_{\rm cont}=\Re[V(r\to\infty)]$ is shown as gray line. A hierarchial destabilization of the bound states in accord with the reduction in binding energy is seen.  Note that depending on the skewness of the peak, the mass it encodes does not have to coincide with it's apex position.  The error bands reflect the uncertainty of the Debye mass determination.
}\label{ET}
\end{figure}

\begin{figure}
\centering
 \includegraphics[scale=0.59]{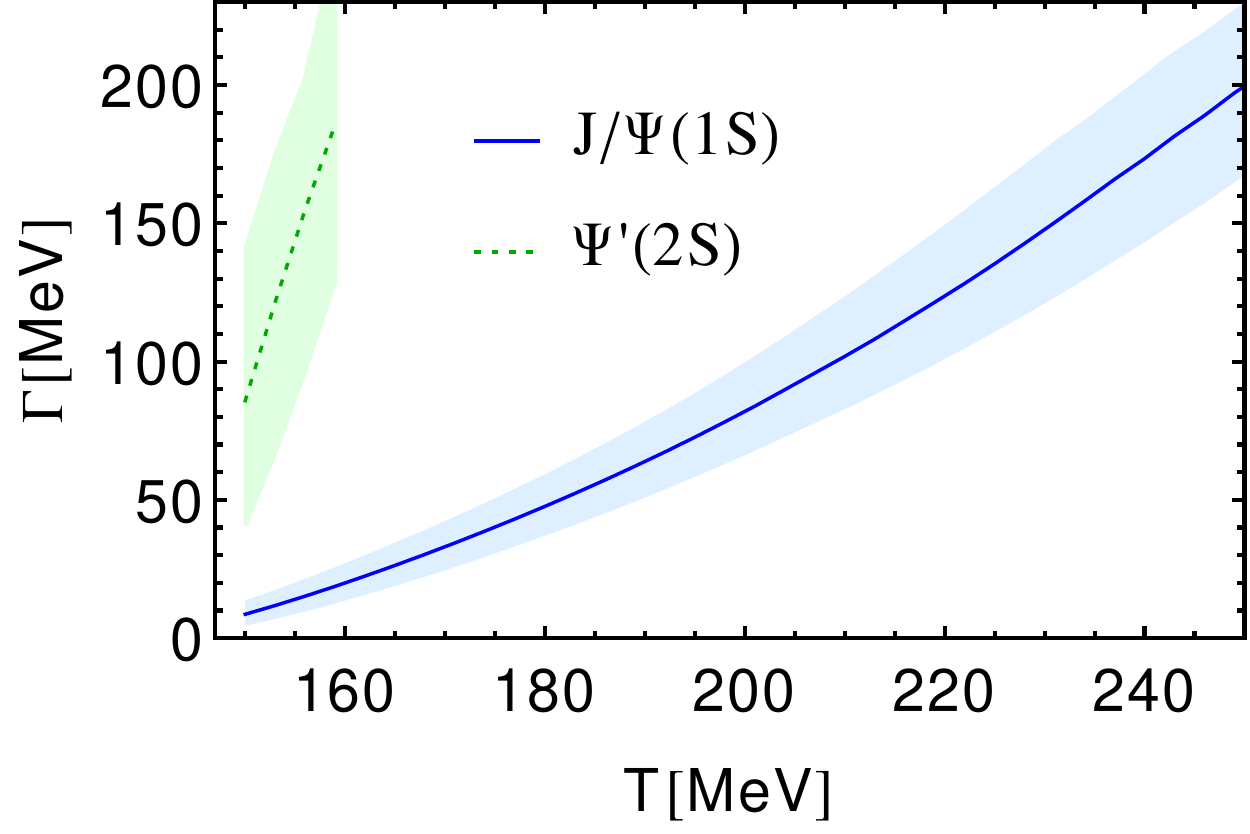}\hspace{0.2cm}
 \includegraphics[scale=0.595]{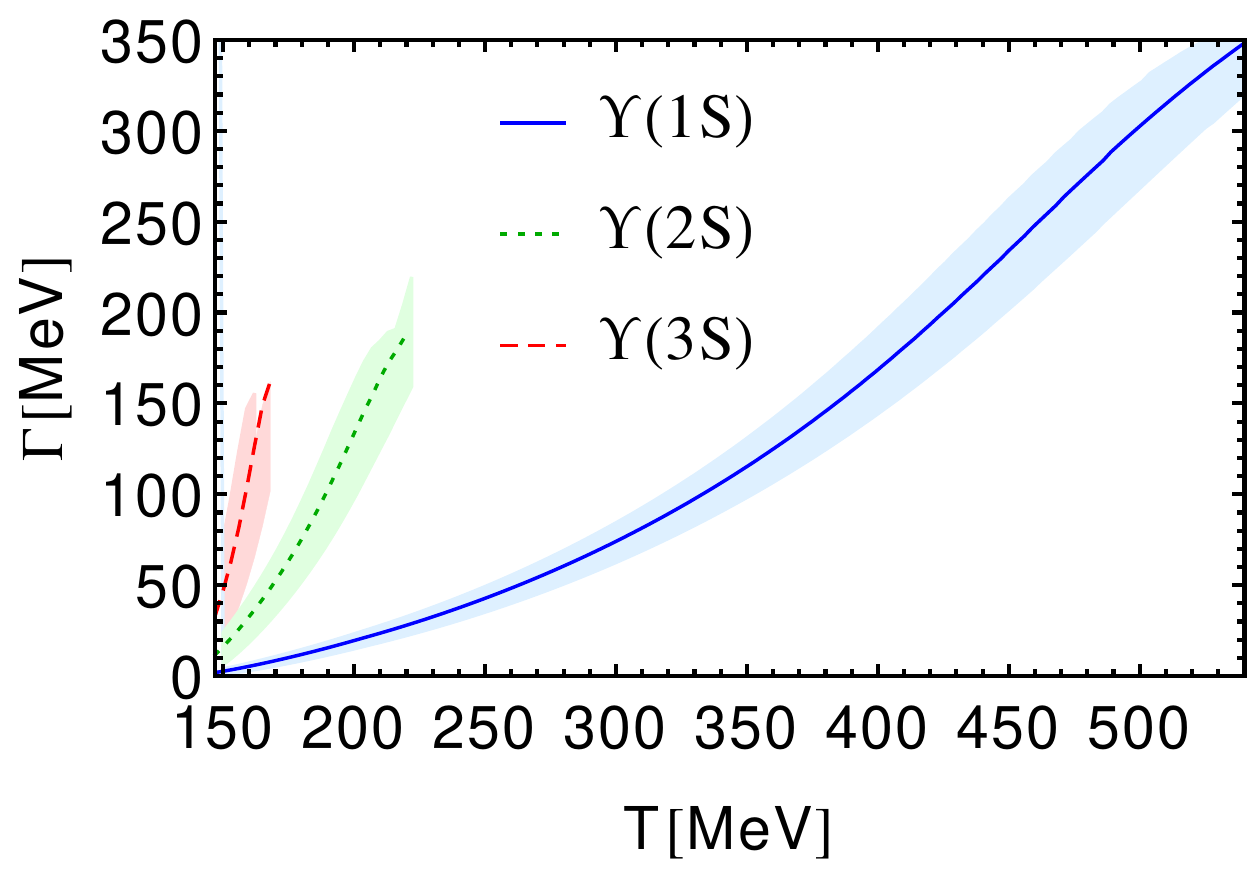}
 \caption{Width of the Charmonium (left) and Bottomonium (right) bound states, which increases monotonously with temperature. The error bands reflect the uncertainty of the Debye mass determination.
}.
\label{WidthT}
\end{figure}

\begin{figure}
\centering
 \includegraphics[scale=0.59]{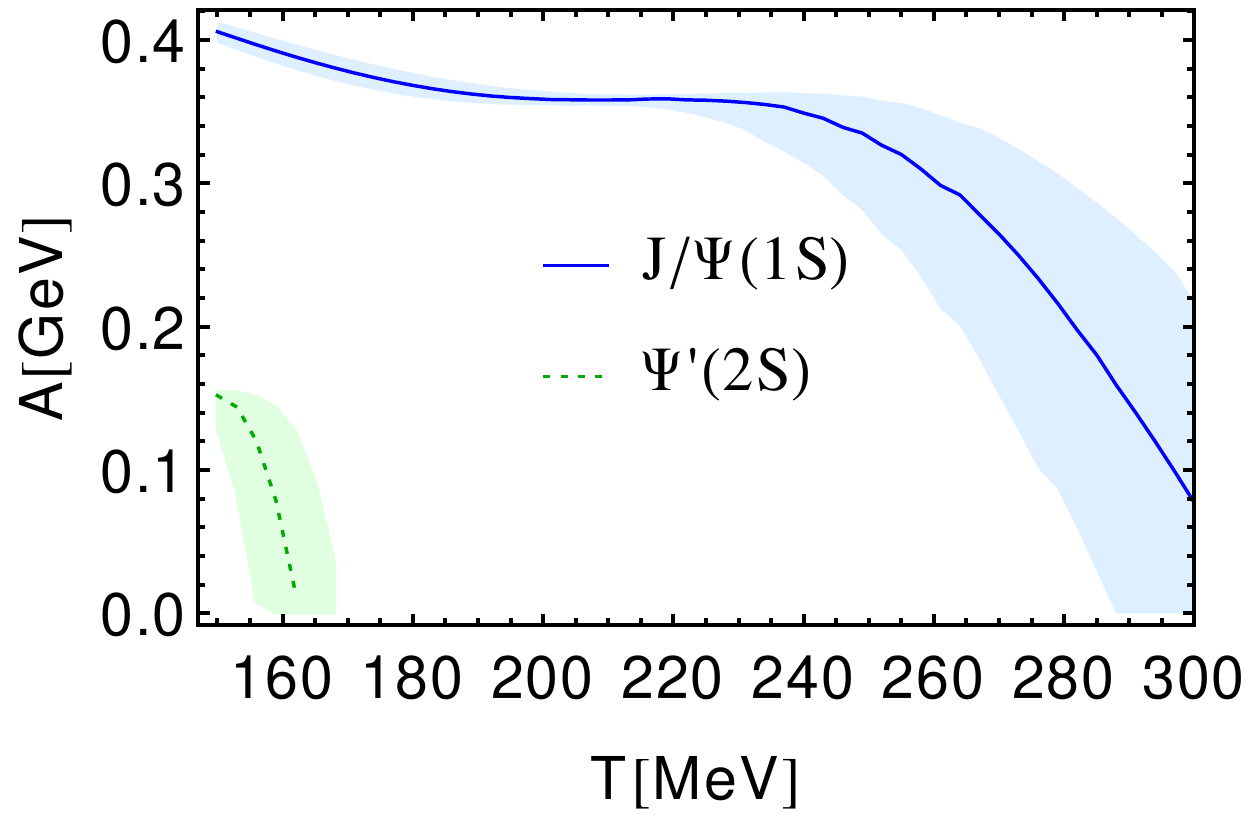}\hspace{0.2cm}
 \includegraphics[scale=0.565]{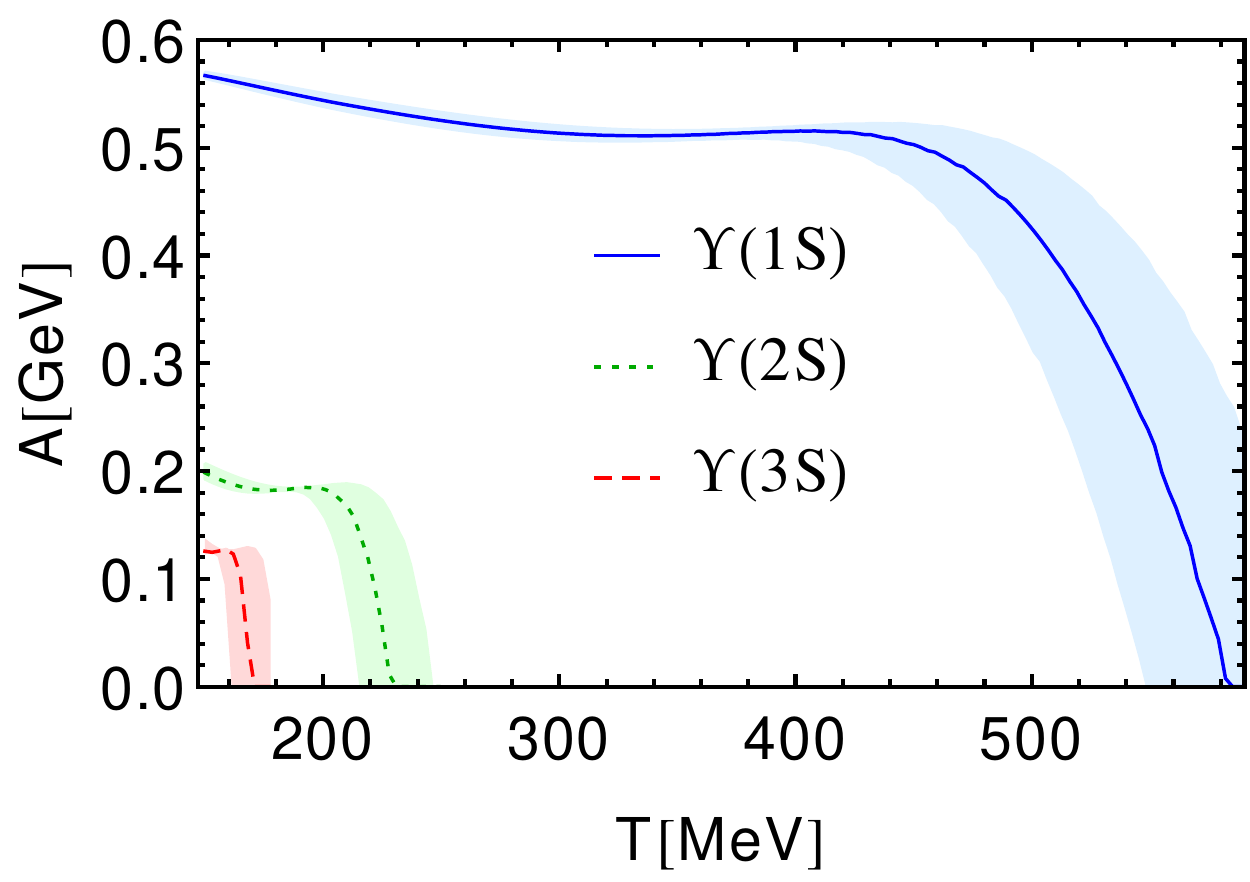}
 \caption{Area under the bound-state peaks in the charmonium (left) and bottomonium (right) spectrum. Note the characteristic plateau region before a rapid decrease to zero sets in. The error bands reflect the uncertainty of the Debye mass determination.}\label{A}
\end{figure}

Using the interpolated form for the Debye mass (\ref{Eq:mD}) we perform a temperature scan of the spectrum at every $\delta_T=3$ MeV and fit the different peaks with the BW of eq.\eqref{sBW}. From these fits we obtain the temperature dependence of the bound state width and mass of different quarkonium states, which are plotted in fig.~\ref{ET} and fig.~\ref{WidthT}  respectively. Another  quantity particularly relevant to phenomenology is the area under each of the peaks, which we define here as the integrated area of the Breit-Wigner fit function $A=\frac{\pi C \Gamma}{2}$ and which is related to the total dilepton emission rate via eq.\eqref{dilepton_prod}. Its values are given in fig.~\ref{A}. We find that as temperature increases, the peak area at first remains rather constant, even if the peak becomes wider but eventually and abruptly begins to decrease rapidly towards zero.

We can put these observations in the context of the in-medium modification of the potential shown in fig.~\ref{Fig:ReImPot}. While the small distance part of the correctly renormalized $\Re[V]$ is virtually temperature independent, a significant screening of the linear rise at large distances occurs with increasing $T$. It is a peculiarity of the confinement mechanism that with the diminishing remnant of the linear rise; also the threshold to the continuum $E_{\rm cont}$ is lowered. This is reflected in a decrease of the value of $E_{\rm cont}=\Re[V(r\to\infty)]$ (see the gray curve in fig.~\ref{ET}). In turn, the binding energy, defined from the difference between bound state mass and the continuum threshold energy is monotonously lowered. That is to say, the thermal fluctuations of the medium destabilize the bound state. Interestingly, once the threshold moves into the vicinity of a formerly firmly bound quarkonium peak it pushes the spectral feature towards lower frequencies until it eventually disappears.

Let us consider the mass shift observed in fig.~\ref{ET}.  At first it might seem counterintuitive, as the expectation for an elementary particle in a medium is exactly the opposite, it will receive a thermal mass. For instance at LO, a single fermion receives a mass correction \cite{Donoghue:1984zz,Seibert:1993aw,Burnier:2014cna} of
\beq
m_Q(T)=m_Q+\delta m_Q^T=m_Q+\frac{g^2 T^2 C_F}{12 m_Q},\label{thermal_shift}
\eeq 
and one might wonder if such a contribution should be added to our potential. As can be seen from \eqref{thermal_shift} it is of higher order in the $1/m_Q$ expansion and hence does not contribute to the static potential. At the next order in the $1/m_Q$ expansion, the potential $V(r)$ is corrected by a term of the form $V_1(r)/m_Q$ \cite{Brambilla:2000gk}. If the two color charges are widely separated one expects that each of them obtains a thermal mass shift, hence $\lim_{r\to\infty} V_1(r)/m_Q=2\delta m_Q^T$. On the other hand, for $r\to 0$ one can imagine the neutral bound state will not interact with the plasma so that $V_1(r)=0$. The thermal mass shift (\ref{thermal_shift}) may hence be considered an upper bound on the thermal mass shift of the bound state. Since it is of higher order it is negligible in our non-relativistic computation, which can be checked using our value of $\tilde\alpha_S = g^2C_F/4\pi$ and considering a temperature of $T\sim 200$MeV. One would obtain $\delta m_Q^T=7$MeV for charm and $\delta m_Q^T=2$MeV for bottom, insignificant in comparison to the thermal shift observed in fig.~\ref{Fig:ChrmBotRhoOvervwS}, which easily reaches $100-300$MeV (see also fig.~\ref{ET}).

These findings, based on a first principles lattice QCD based complex in-medium potential, are qualitatively similar to what had been observed in potential modeling studies that solve a Schr\"odinger equation with a potential with imaginary part inserted by hand (see e.g. \cite{Petreczky:2010tk}). There only the Coulombic contribution to ${\rm Im}[V]$ was used, which leads to more stable behavior than in our case. The presence of the string-like vacuum potential contributes an additional term to ${\rm Im}[V]$, which is of comparable size as the Coulombic contribution to ${\rm Im}[V]$ at intermediate temperatures. On the other hand our results differ significantly from the spectra obtained in the T-matrix study of ref.~\cite{Riek:2010fk}. There the authors use purely real model potentials which lead to in-medium states that appear to possess a significantly larger energy than their vacuum counterparts. Updated computations in that framework \cite{Liu:2015ypa} are expected in the near future. Similar shifts of the resonance peaks to lower energies were also observed in a sum-rule based approach \cite{Dominguez:2009mk,Dominguez:2010mx,Dominguez:2013fca}, but the magnitude of the shift is somewhat weaker there. A more thorough comparison of our results to direct studies in lattice QCD is part of ongoing work.

For $T\sim250$MeV one can also compare to the perturbative estimates of \cite{Burnier:2008ia} and qualitatively good agreement between the spectra is found. The lattice spectra show slightly narrower peaks than the perturbative ones, due to the linearly rising part of the potential, which increase the binding energy. This effect is not captured in perturbation theory. On the other hand, the string effects vanishes quickly at high temperature and in addition the imaginary part of the lattice potential is larger due to string effects. Combined it seems to compensate the stronger binding through an increase in the width of the state. One central benefit of the lattice computation is the possibility to connect the high and low temperature spectra, which is cannot be realized in perturbation theory.

A qualitative difference between the perturbative results and the lattice ones is the way the peak positions change as function of the temperature. In perturbation theory, which contains essentially only the Coulomb term, the bound state peaks move slightly to higher values of frequency \cite{Burnier:2007qm} as T increases whereas we see here a clear decreases of the peak frequencies.

\begin{figure}[t]
\centering
 \includegraphics[scale=0.59]{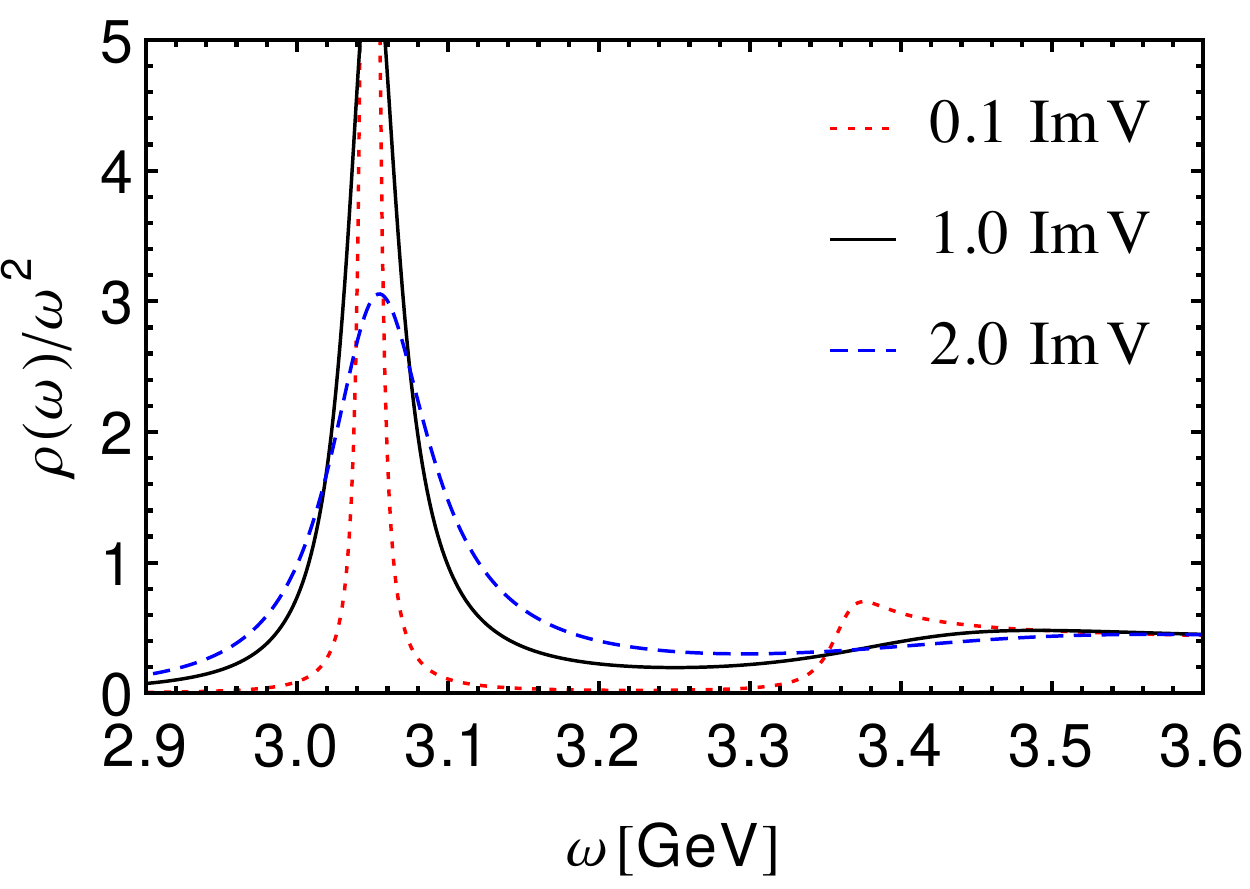}\hspace{0.2cm}
 \includegraphics[scale=0.575]{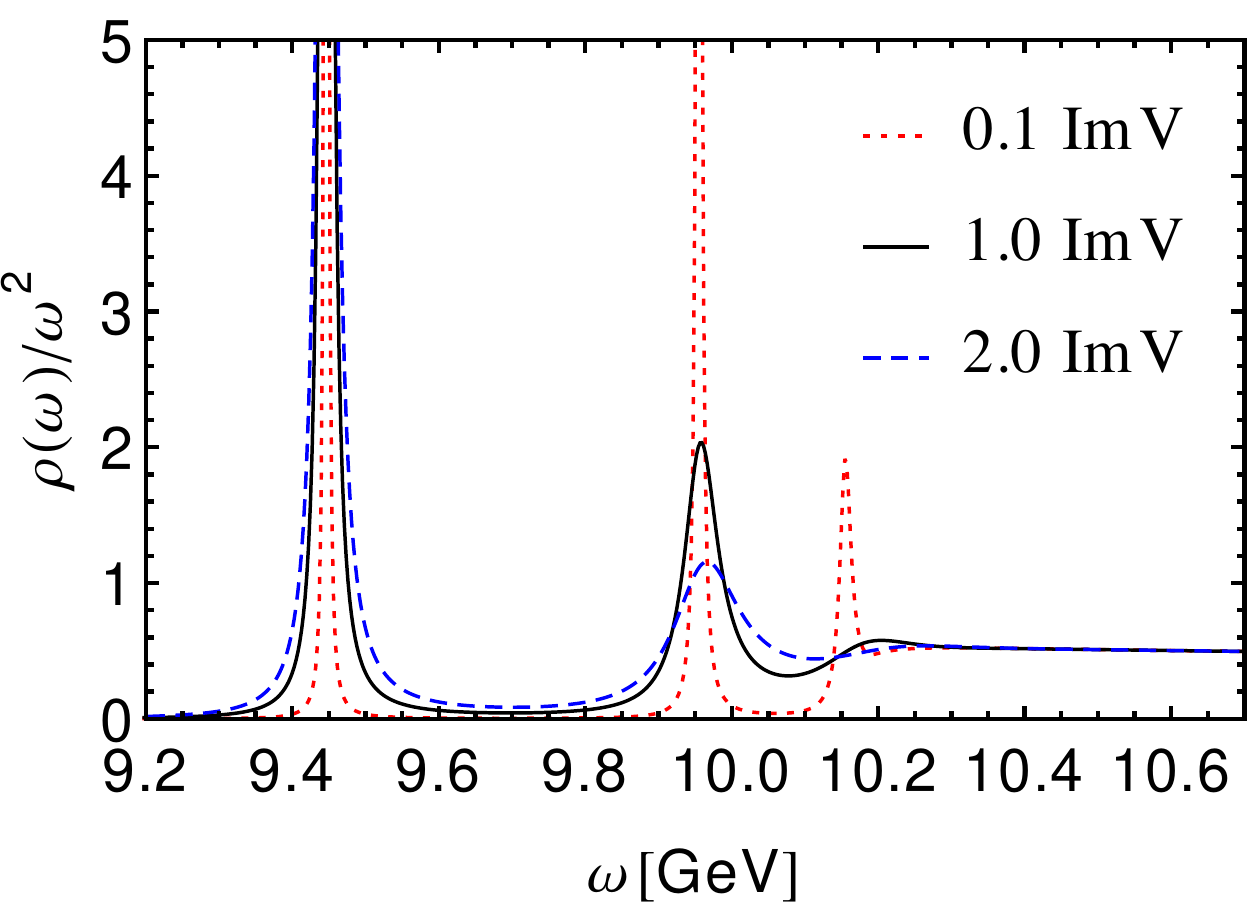}
 \caption{Charmonium (left) and Bottomonium (right) spectrum at $T=1.19T_C$ with manually changed values for the imaginary part of the potential ($0.1\,\Im V$ (black), $\Im V$ (red) and $2\,\Im V$ (blue)). While the peak position hardly changes unless the peak is close to melting, the width significantly depends on the strength of $\Im V$).
 }\label{ImV_test}
\end{figure}

As the determination of the imaginary part of the potential represents a significant source of uncertainty in this work, we close this section by studying its effect on the spectrum in more detail. To do so, we vary the strength of the imaginary part multiplying it by a $r$-independent factor see fig.~\ref{ImV_test}. The main effect of the imaginary part is to broaden the peak, without changing its position and area, unless the peak is already close to the continuum, as is e.g. the case with $\Upsilon(3S)$ in fig.~\ref{ImV_test}. That said, it is obvious that the corresponding states do melt more quickly with a large ${\rm Im}[V]$, as the width reaches the binding energy much more quickly. 

\FloatBarrier

\section{In-medium quarkonium penomenology}

\subsection{Charmonium at freezout}
\label{sec:charmonium}

At current heavy-ion colliders, such as RHIC and LHC, with collision energies above $\sqrt{s_{AA}}>200$GeV, the success of the statistical model of hadronization, to predict the yields of heavy quarkonium, supports the idea that charmonium completely dissolves in the created plasma. In turn essentially all charmonium bound states we observe in experiment would be generated via recombination that takes place at the freeze-out boundary, usually located slightly below the crossover temperature. In such a setting the ratio between the yields for $J/\Psi$ and $\Psi(2S)$ can be estimated from the difference in area under the corresponding peak structures in the in-medium spectral functions in a straight forward way. 

\begin{figure}
\centering
\includegraphics[scale=0.6]{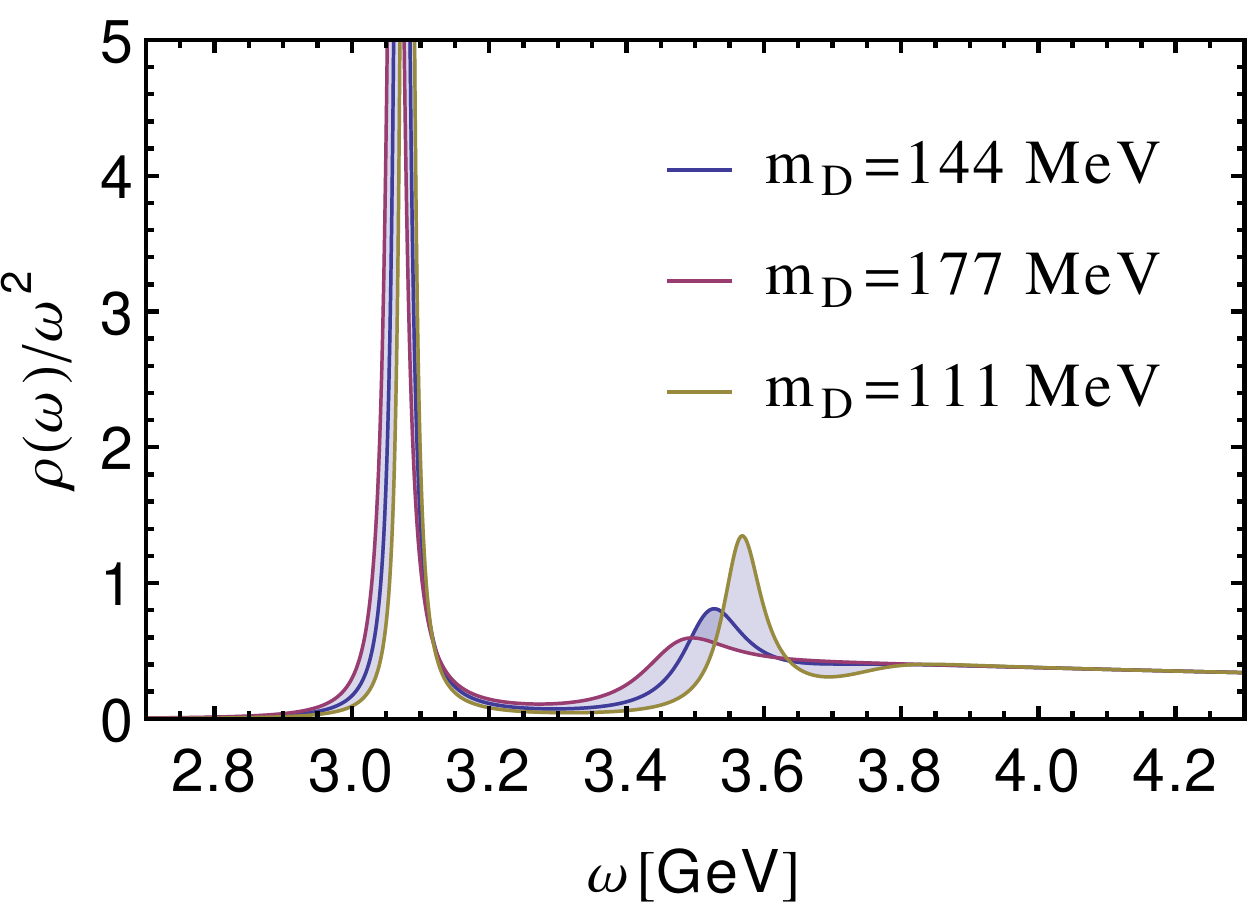}
\includegraphics[scale=0.6]{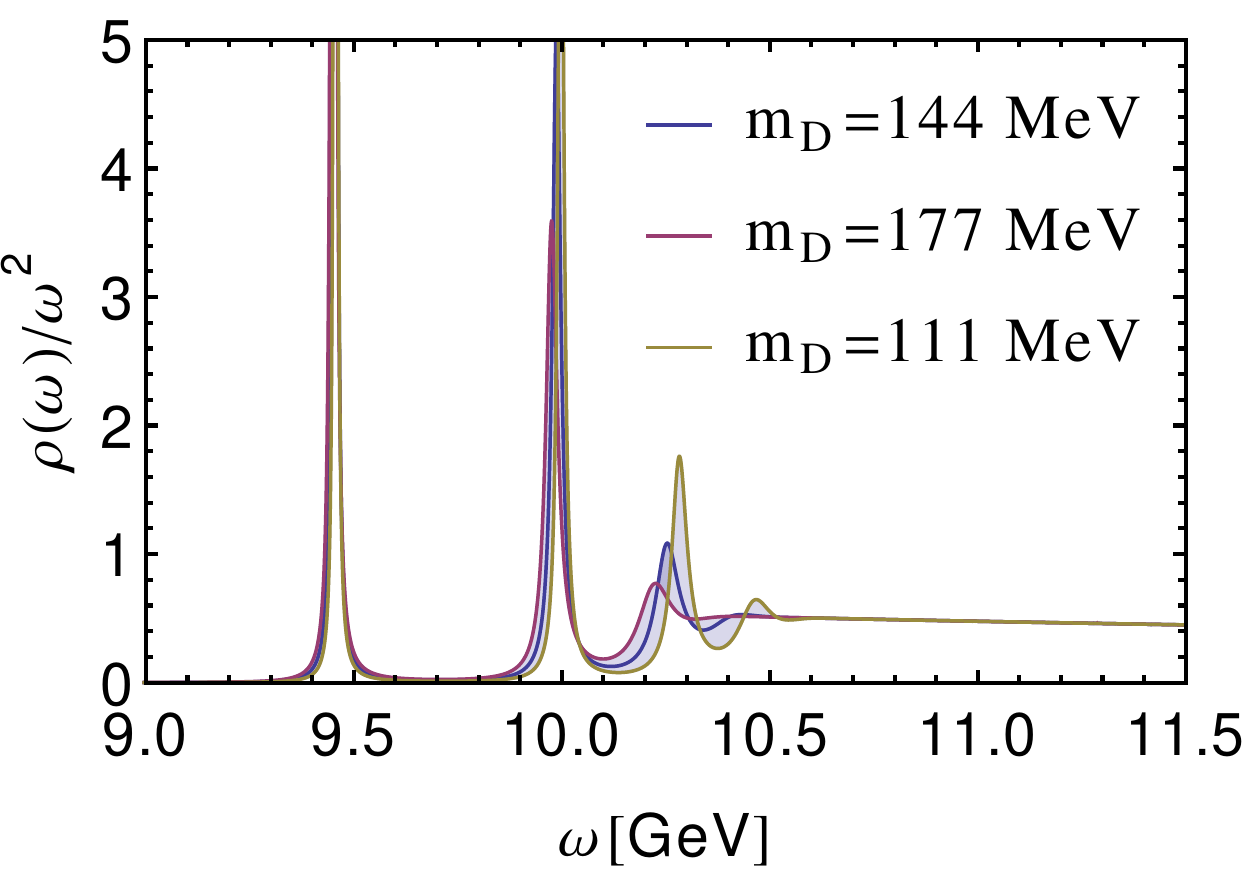}
\vspace{-0.2cm}
\caption{charmonium (left) and bottomonium (right)  spectra at $T_c$. To visualize the uncertainties of the computation three curves according to the values of $m_D$ and $m_D\pm\delta m_D$ are shown. } 
\label{fig::charm_at_freezout}
\end{figure}

Let us assume that due to the dynamical nature of the expanding fireball the freeze-out happens close to the chiral crossover temperature $T_C$.  When inspecting the spectra at this temperature, as done in fig.~\ref{fig::charm_at_freezout} we find that there indeed exist two peaked structure related to the in-medium $J/\Psi$ and $\Psi'$. I.e. we can compute the amount of in-medium dimuons produced by each of the two states via the product of the spectral function and Bose-Einstein distribution, integrated over the frequency range corresponding to each of the states via eq.~(\ref{dilepton_prod}).

This however is not what is measured in experiment, since the charmonium states do not  decay into leptons within the plasma but given their life-time, instead long after the plasma is diluted away. We should thus in principle project the states corresponding to the peaks of the finite $T$ spectra onto the $T=0$ states. As no agreed upon method exists to do so, we here assume that the states inside the peaked structures will become real $J/\Psi$ or $\Psi'$ after freeze out and subsequently decay outside of the plasma.
These assumptions are similar to those made in the statistical model \cite{Andronic:2009sv} and take into account the in-medium modification of the particle states. 

In order then to calculate the phenomenologically relevant density of charmonium states at freeze out from the spectral function, we use the following tactic: We estimate first the contribution from the different states to the in-medium dilepton emission rate. From it we can compute the ratio of dileptons produced by $\Psi'$ and $J/\Psi$. The ratio of the number densities of $\Psi'$ and $J/\Psi$ follows when we correct for the probability of the corresponding vacuum states to decay electromagnetically.

In detail, the dilepton emission rate (\ref{dilepton_prod}) reads
\beq
R_{\ell\bar\ell}\propto \int dp_0 d^3\bp \frac{\rho(P)}{P^2}n_B(p_0).
\eeq
Here we use the fact that to leading order $\rho$ depends only on $P^2=p_0^2-\bp^2$ and leave a more detailed analysis of the momentum dependence for future studies. After performing the change of variable $\omega=\sqrt{p_0^2-\bp^2}$, we obtain
\beq
R_{\ell\bar\ell}\propto \int d\omega\, d^3\bp\, \frac{\rho(\omega)}{\omega^2} n_B(\sqrt{\omega^2+\bp^2})\frac{\omega}{\sqrt{\omega^2+\bp^2}}\label{int}.
\eeq
In this formula, the contribution from the different bound states will come from the corresponding peak area in $\rho(\omega)/\omega^2$, see fig.~\ref{fig::charm_at_freezout}. Hence we fit $\rho(\omega)/\omega^2$ with a skewed Breit-Wigner (\ref{sBW}) to distinguish the contribution from the different states $n$. From that fit we obtain the position and width of the spectral feature, i.e. the thermal mass $M_n$ of the particle and its width.

\begin{figure}
\centering
\includegraphics[scale=0.6]{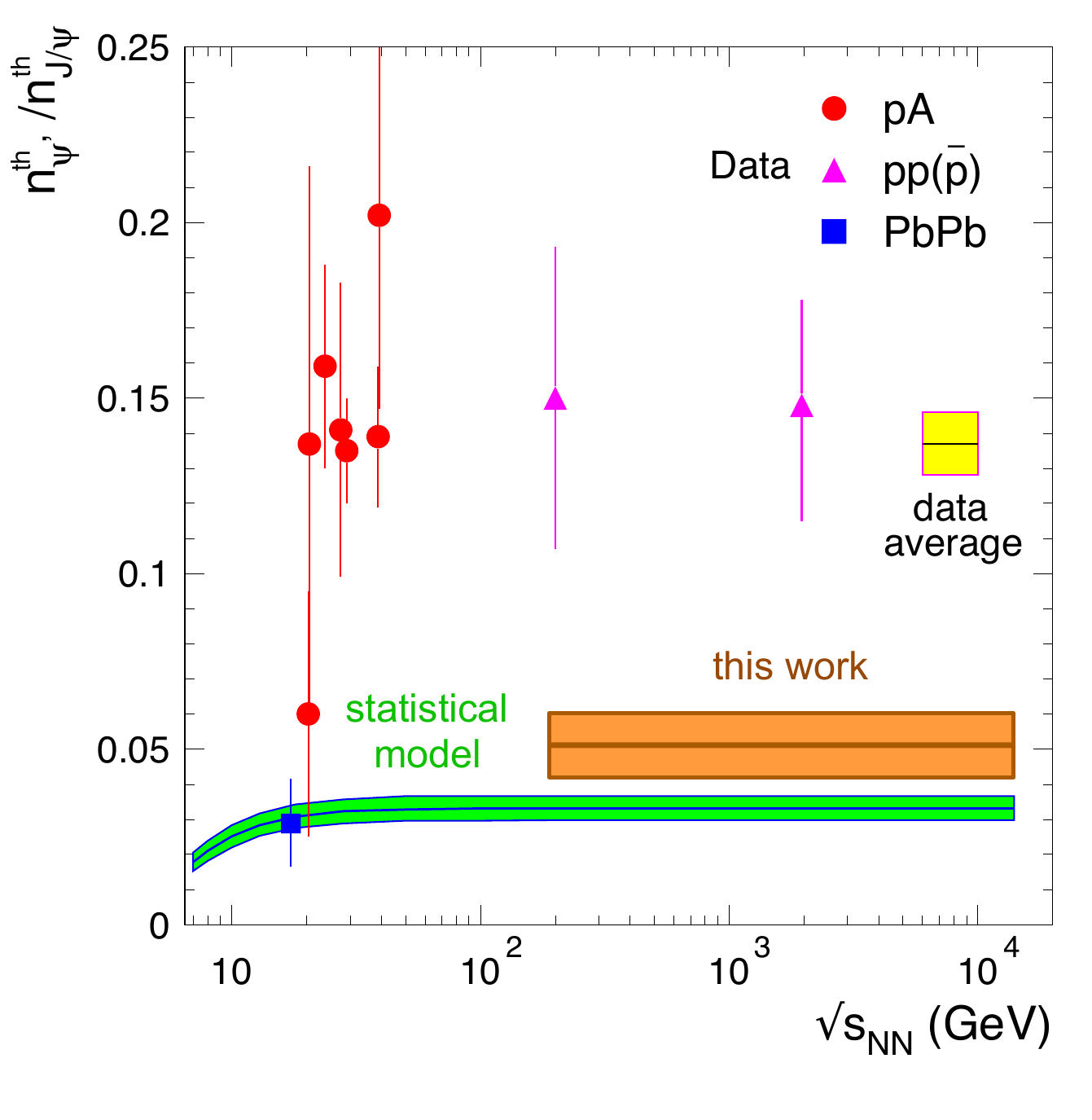}
\vspace{-0.2cm}
\caption{Comparison of predicted yields for $\Psi'$ and $J/\Psi$ from this work (orange) compared to predictions from the statistical model (green) and several experimental measurements. The plot structure and experimental data has been adapted from ref.~\cite{Andronic:2009sv}. } 
\label{Fig:StatisticalModelComp}
\end{figure}

To calculate the integral (\ref{int}) we checked numerically that it is possible to approximate the Breit-Wigner peak by a delta peak, keeping the area $A$ under the curve constant. Performing this, the integral (\ref{int}) becomes
\beq
R_{\ell\bar\ell}^{\Psi_n} \propto A \int d\omega\, d^3\bp\, n_B(\sqrt{M_n^2+\bp^2})\frac{M_n}{\sqrt{M_n^2+\bp^2}}.
\eeq
The fit of the charmonium spectrum in fig.~\ref{fig::charm_at_freezout} then yields
\beq
\frac{R_{\ell\bar\ell}^{\Psi'}}{ R_{\ell\bar\ell}^{J/\Psi}}=0.023\pm0.004.
\eeq
To obtain the number density we divide, as discussed, by the electromagnetic decay rate of a vacuum state, which is proportional to the square of the wave function $\Phi$ at $r=0$ divided by the square of the mass of the state \cite{Bodwin:1994jh}
\beq
\left. \frac{N_{\Psi'}}{ N_{J/\Psi}} \right|_{T=T_C}=\frac{R_{\ell\bar\ell}^{\Psi'}}{ R_{\ell\bar\ell}^{J/\Psi}} \frac{ M_{\Psi'}^2 |\Phi_{J/\Psi}(0)|^2}{M_{J/\Psi}^2 |\Phi_{\Psi'}(0)|^2} =0.052\pm0.009,
\eeq
where we used the value at the origin of the zero temperature wave functions. A comparison of our result to those of the statistical model is shown in fig.~\ref{Fig:StatisticalModelComp}.

While the experimental determination of the $\Psi'$ to $J/\Psi$ ratio is challenging and currently limited by statistics, both ALICE \cite{Adam:2015isa} and CMS \cite{Khachatryan:2014bva} have put forward first measurements of the double ratio $\left. \frac{N_{\Psi'}}{ N_{J/\Psi}}\right|_{\rm PbPb}/\left. \frac{N_{\Psi'}}{ N_{J/\Psi}}\right|_{\rm pp}$. The cancellation of experimental systematic uncertainties makes this quantity the most robust measure of the in-medium relation between $\Psi'$ and $J/\Psi$ to date. Integrated over centrality CMS found that at the lowest accessible rapidities the ratio lies at $0.45\pm0.13 {\rm (stat)}\pm0.07 {\rm (syst)}$  , while at larger rapidity  $1.6<|y|<2.4$ it grows to values larger than one.

Here we give a rough estimate of the double ratio by using experimental data obtained for prompt charmonium at $\sqrt{s}=7$TeV by the CMS collaboration. The rapidity averaged cross-section ratios for $\Psi'$ to $J/\Psi$ denoted by $R$ in \cite{Chatrchyan:2011kc} are furthermore averaged over transverse momentum $\langle R\rangle_{p_T}=0.0378\pm0.006$, as the observed dependence on $p_T$ is rather weak. To extract the number density ratio, the difference in branching fraction into dimuons must also be corrected, after which we obtain
\beq
\left. \frac{N_{\Psi'}}{ N_{J/\Psi}} \right|_{\sqrt{s}=7{\rm TeV}}=\langle R\rangle_{p_T}\,\frac{BR ( J/\Psi\to \mu^+\mu^-)}{BR ( \Psi'\to \mu^+\mu^-)}  =0.09\pm0.015,
\eeq
This leads to a double ratio of 
\beq
\left. \frac{N_{\Psi'}}{ N_{J/\Psi}}\right|_{\rm PbPb}/\left. \frac{N_{\Psi'}}{ N_{J/\Psi}}\right|_{\rm pp} = 0.58\pm 0.14
\eeq
in which the errors in the individual contributions have been naively propagated under the assumption of being independent. Compared to the low rapidity measurements by CMS \cite{Khachatryan:2014bva} we find good agreement within our still sizable error bars. 

\FloatBarrier

\subsection{Bottomonium at maximum LHC temperatures}
\label{sec:botphen}

\begin{figure}[t!]
\centering
\includegraphics[scale=0.5725]{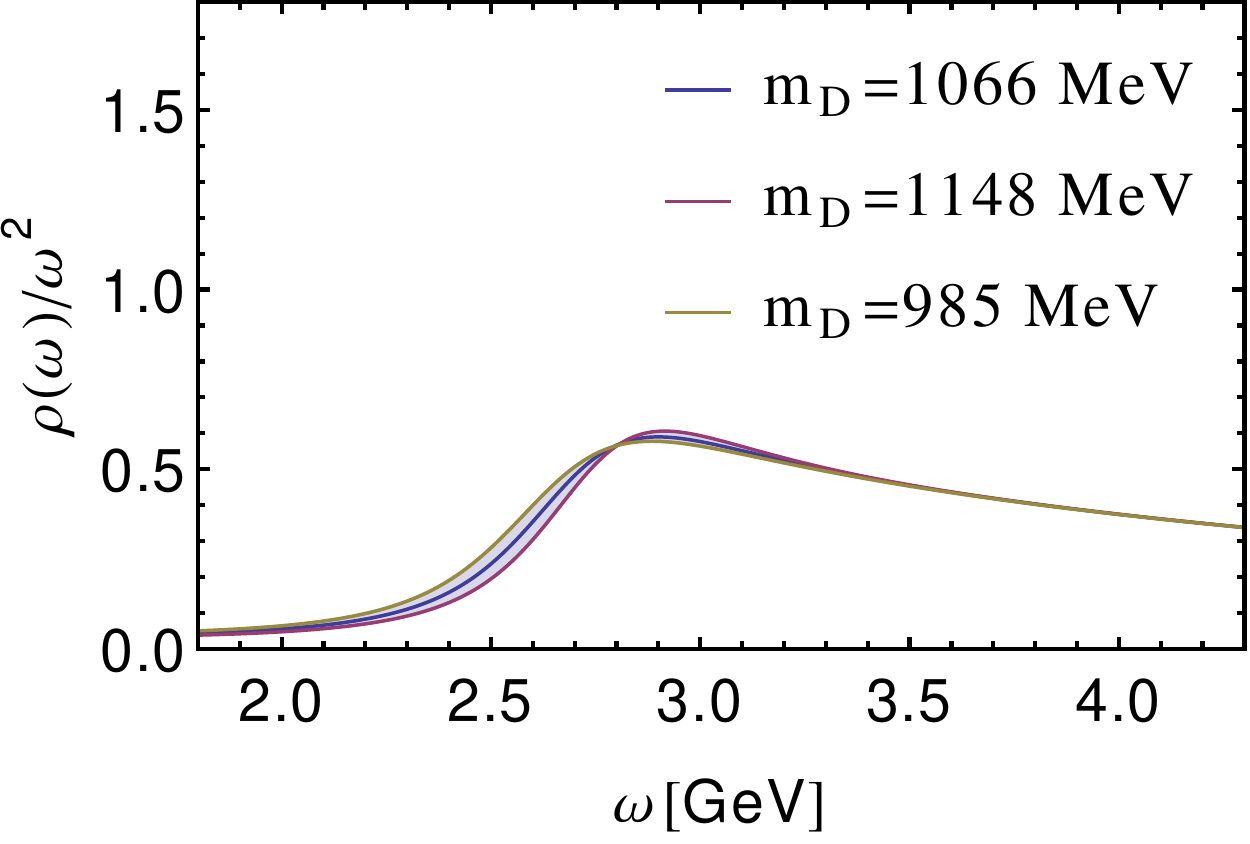}
\includegraphics[scale=0.6]{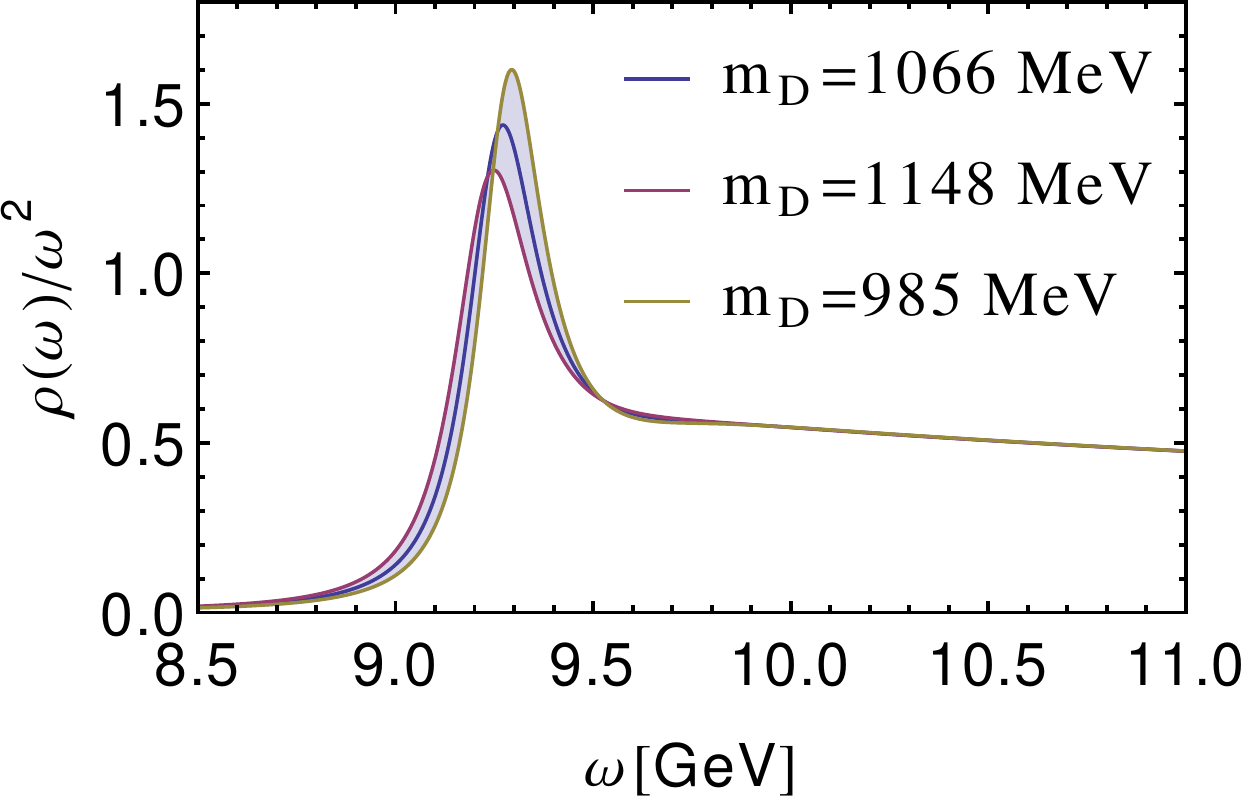}
\vspace{-0.2cm}
\caption{charmonium (left) bottomonium (right) spectra at $T=3 T_C$ from an extrapolation of the HTL fitted values for $m_D$. To visualize the uncertainties of the computation three curves according to the values of $m_D$ and $m_D\pm\delta m_D$ are shown.} 
\label{fig::bottom_at_Tmax}
\end{figure}

Given the small amount of bottom quarks produced in heavy-ion collisions, both at RHIC
and LHC, one would naively expect a negligible amount of recombination, a thought 
which might however need reconsideration \cite{Young:2008he,Young:2009tj}.
The bottomonium states seen in lead-lead collisions are hence expected to be
the ones that survive over the whole plasma evolution. A
way to approximate their number is to read off the peak area from the spectrum at the highest
temperature reached by the plasma. At LHC, where most experimental results for bottomonium 
have been obtained, the highest temperature according to ref.~\cite{Habich:2014jna} corresponds to $T\simeq3T_C$ from an analysis of harmonic yields.
We hence consider the spectrum at $T_{\rm max}=3T_C$ in fig.~\ref{fig::bottom_at_Tmax}.
There, we see that only the ground state survives and is noticeably
washed out.
Of course not all the bottomonium formed will experience the highest
temperature, one e.g. expects states to be also produced at the edge of
the plasma, where the temperature is lower.

If we assume the fate of the heavy quarkonium is determined at the maximum temperature reached by the
plasma, we can estimate the ratio of bottomonium produced in lead-lead
collisions to the amount of Bottom produced in proton-proton collisions, which is given by the ratio of the spectral weights at $T_{\rm max}=3T_C$ vs $T=0$,
\beq
R_{\rm PbPb/pp}=\frac{\Gamma_{T_{\rm max}} C_{T_{\rm max}}}{\Gamma_{T=0} C_{T=0}}=0.91\pm0.02.
\eeq
Here $\Gamma$ denotes the width and $C$ the amplitude of the corresponding Breit-Wigner type spectral feature. Note that the error bars might be underestimated as the Debye mass had to be extrapolated far above the available temperatures using the HTL fit. The corresponding experimental results from CMS \cite{Hu:2011zza} are lower with $R_{\rm PbPb/pp}=0.6\pm0.2$.

\subsection{Melting temperatures}

As we found in fig.~\ref{Fig:ChrmBotRhoOvervwS}, the bound states disappear progressively with increasing $T$ and we can attempt to define their melting point. In the case of a real potential, the melting temperature is straightforward to define from the disappearance of the purely real binding energy. For a complex potential the situation is more subtle, as the bound state broadens before it disappears. A popular choice is to define the melting of a state the moment its width equals its binding energy \cite{Laine:2007qy,Kakade:2015xua}. 

Here we scan the spectrum at different temperatures, using the HTL based inter- and extrapolation of the values of the continuum corrected Debye mass $m_D$. The bound state peak features are determined from a fit based on the skewed Lorentzian of eq.(\ref{sBW}). The melting temperatures obtained in this way are given in tab.~\ref{t::melting} and exhibit a hierarchical pattern, where only the ground states survive well above $T_C$\footnote{For charmonium a similar behavior has been anticipated in the in the sequential melting scenario of ref.~\cite{Karsch:2005nk}.}.

Our observations are in qualitative agreement with earlier lattice QCD studies of charmonium correlators and spectral functions 
\cite{Ding:2012sp,Asakawa:2003re,Datta:2003ww,Jakovac:2006sf,Iida:2006mv,Ohno:2011zc,Aarts:2007pk,Borsanyi:2014vka,Ohno:2014uga}. Note that most of these studies were performed in the quenched approximation or at
rather large values of the light quark masses and so far no continuum extrapolation was performed.
Results of spatial correlation functions and the corresponding spatial screening masses from lattice calculations in the charmoniun sector that include dynamical light quarks further support our findings \cite{Bazavov:2014cta,Karsch:2012na}.
Relativistic charmonium and bottomonium correlation functions computed in quenched
QCD \cite{Ohno:2014uga} have also shown that in the bottomonium channel thermal
modifications set in at larger temperatures, well in the QGP phase, compared to
the charmonium channel. The recent determination of bottomonium spectra in full QCD based
on non-relativistic QCD \cite{Aarts:2014cda,Kim:2014iga} corroborates an $\Upsilon(1S)$ ground state survival deep into the QGP. A thorough quantitative comparison of our lattice potential based spectra to direct lattice QCD computations will be part of a future study and is work in progress.

\begin{table}[h!]
\centering
 \begin{tabular}{|c|c|c||c|c|c|c|}\hline
states& $ J/\Psi(1S)$& $\Psi'(2S)$ & $\Upsilon (1S) $ &$\Upsilon (2S) $ &$\Upsilon (3S) $ &$\Upsilon (4S) $ \\ \hline\hline
 $T_{\rm melt}^{\Gamma=E_{\rm bind}}/T_C $ &$1.37_{-0.07}^{+0.08}$& $<0.95$  & $2.66_{-0.14}^{+0.49}$ & $1.25_{-0.05}^{+0.17}$ & $ 1.01_{-0.03}^{+0.03}$ & $<0.95$\\   \hline
 \end{tabular}
 \caption{Melting temperatures $T_{\rm melt}$ of the different bound states defined by the point, at which the width of the state equates its binding energy calculated from $\Re[V]$ alone. The error on the determination of $T_{\rm melt}$ takes into account the possible variation of $m_D$ as shown in fig.~\ref{Fig:mDFits}. Note that because of the lack of sufficient number of lattice ensembles below $T=0.95T_C$ and the breakdown of the generalized Gauss law ansatz, we only give upper limits in this regime.}\label{t::melting}
\end{table}

A more precise determination of the $\Upsilon (1S) $ melting temperatures will require lattice simulations at higher temperatures, as for now we can only use an extrapolation of the HTL fit for $m_D$, which becomes increasingly unreliable at high temperatures beyond $T>1.66T_C$.

\FloatBarrier

\section{Conclusion}
\label{sec:concl}

Heavy quarkonium is a vital probe to uncover the physics of the quark-gluon plasma created in heavy-ion collisions. Their non-relativistic nature opens up the possibility to describe their in-medium behavior by an effective potential entering a Schr\"odinger equation. We reported here the first quantitative phenomenological investigation of bottomonium and charmonium S-wave spectra at finite $T$, based on the proper complex in-medium static potential extracted from $N_f=2+1$ lattice QCD.

To make possible the use of the lattice potential in a phenomenological application we deployed the generalized Gauss law ansatz, which provides analytic expressions for $\Re[V]$ and $\Im[V]$ that depend on a single temperature dependent parameter $m_D$, the Debye mass. Tuning $m_D$ it was possible to reproduce the lattice values of $\Re[V]$ and to validate the up to now tentative values of $\Im[V]$ at all separation distances and temperatures investigated. After correcting for lattice discretization artifacts in the Debye mass we fitted its temperature dependence successfully with a HTL based expression. 

To compute spectral functions for phenomenological inspection, we determined the vacuum parameters of the static potential in the continuum from a fit to the experimentally known S-, P- and D-wave bottomonium states. Finite $T$ effects are incorporated though the continuum corrected Debye mass, extracted from the real-part of the in-medium lattice potential, leading to correctly renormalized finite temperature expressions for $\Re[V]$ and $\Im[V]$

The central findings in our adiabatic setup are a clear pattern of sequential melting of both bottomonium and charmonium with respect to their vacuum binding energies. The interplay between Debye screening and scattering with medium partons leads to characteristic shifts of the in-medium bound states to lower masses before they dissolve into the continuum. This effect is opposite to the relatively small gain in thermal mass.

The availability of in-medium spectra with physical widths allowed us to estimate the $\Psi'$ to $J/\Psi$ ratio at the crossover transition temperature, relevant in the context of the proposed statistical hadronization scenario. Our approach to compute the ratio of in-medium dimuon emission corrected by the emission rate of the corresponding vacuum states yields a value slightly larger than that of the statistical model but still much smaller than the experimental data in $p+p$ collisions. A rough estimate of the suppression of the $\Upsilon(1S)$ state in this adiabatic scenario on the other hand gave a value $50\%$ larger than experimentally observed.

The static potential is only the starting point for a quantitatively reliable investigation of heavy quarkonium. In particular for the lighter flavor charmonium the question of finite mass effects in the potential should be addressed and efforts should be directed at determining the first correction $V_1(r)/m_Q$ at finite temperature from the lattice. On the other hand direct lattice QCD studies, be it in the relativistic formulation or using the effective field theory NRQCD including velocity correction, will be essential to crosscheck the validity of the potential description. Comparisons with the data here can be made both on the level of spectra, as well as the corresponding Euclidean correlators. The necessity to solve an inherently ill-defined problem in order to extract the spectral functions from lattice QCD current-current correlators directly remains a significant challenge. In order to surmount it both methodological progress in Bayesian and non-Bayesian spectral reconstruction approaches, as well as computational efforts to generate lattices with more Euclidean time steps will be required.

The lattice in-medium potential can also be used to setup a description of the real-time evolution of the heavy quarkonium wave function at finite temperature in the context of open quantum systems. An approach based on a stochastic potential \cite{Akamatsu:2011se} appears promising, where a purely real potential, given by $\Re[V]$ is randomly perturbed at each time step by noise of strength related to $\Im[V]$. If developed further it promises to provide vital and phenomenologically relevant information on e.g. the density matrix of states for the quarkonium system, which goes beyond what is accessible from within the spectral functions computed here.

Directions for future work include the determination of the complex in-medium potential on lattices at higher temperature in order to avoid the extrapolations beyond $T=1.66T_C$ required e.g. in the determination the $\Upsilon(1S)$ melting temperature in sec.~\ref{sec:botphen}. Another aspect of interest is the momentum dependence of the spectra which is also an active area of research in direct lattice QCD studies. 

\section*{Acknowledgments}

The authors thank G.~Aarts, A.~Andronic, N.~Brambilla, M.~Laine, P.~Petreczky and B.~Keren-Zur for stimulating discussions. We furthermore thank W. Soeldner and the Bielefeld-BNL-CCNU Collaboration for providing the asqtad gaugefield configurations used in this study. Part of the calculations were performed on the Bielefeld GPU cluster and on the ITP in-house cluster in Heidelberg. YB is supported by SNF grant PZ00P2-142524.

\end{document}